\documentclass[preprint,authoryear]{elsarticle}
\usepackage[T1]{fontenc}
\usepackage[margin=1.5in]{geometry}
\usepackage{graphicx} % Required for inserting images
\usepackage{amssymb}
\usepackage{amsmath}
\usepackage{amsfonts}
\usepackage{multirow}
\usepackage{tabularray}
\usepackage{natbib}
\setcitestyle{numbers,square}
\usepackage{url}

\usepackage{booktabs}
\usepackage[usenames,dvipsnames]{xcolor}
\definecolor{gray}{rgb}{0.7,0.7,0.7}
\definecolor{dblue}{rgb}{0,0,0.75}
\definecolor{dred}{rgb}{0.6,0,0}
\definecolor{dgreen}{rgb}{0,0.5,0}

\usepackage{tikz}
\usetikzlibrary{positioning}
\usetikzlibrary{patterns}
\usetikzlibrary{arrows.meta}

\usepackage{tabularray}

\makeatletter
\def\ps@pprintTitle{%
    \let\@oddhead\@empty
    \let\@evenhead\@empty
    \def\@oddfoot{\footnotesize\itshape
         {Submitted preprint} \hfill\today}%
    \let\@evenfoot\@oddfoot
    }
\makeatother

\begin{document}

\title{Multiple split approach -- multidimensional probabilistic forecasting of electricity markets}

\author[1]{Katarzyna Maciejowska}
\ead{katarzyna.maciejowska@pwr.edu.pl}

\author[1]{Weronika Nitka\corref{cor1}}
\ead{weronika.nitka@pwr.edu.pl}

\affiliation[1]{organization={Department of Operations Research and Business Intelligence, Wrocław University of Science and Technology}, 
addressline={Wybrzeże Wyspiańskiego 27}, 
postcode={50-370},
city={Wrocław}, 
country={Poland}}

\cortext[cor1]{Corresponding author}


\maketitle

%-----------------------------------------------------------------
\section*{Abstract}

In this article, a \textit{multiple split} method is proposed that enables construction of multidimensional probabilistic forecasts of a selected set of variables. The method uses repeated resampling to estimate uncertainty of simultaneous multivariate predictions. This nonparametric approach links the gap between point and probabilistic predictions and can be combined with different point forecasting methods. The performance of the method is evaluated with data describing the German short-term electricity market. The results show that the proposed approach provides highly accurate predictions. The gains from multidimensional forecasting are the largest when functions of variables, such as price spread or residual load, are considered.

Finally, the method is used to support a decision process of a moderate generation utility that produces electricity from wind energy and sells it on either a day-ahead or an intraday market. The company makes decisions under high uncertainty because it knows neither the future production level nor the prices. We show that joint forecasting of both market prices and fundamentals can be used to predict the distribution of a profit, and hence helps to design a strategy that balances a level of income and a trading risk. 

%-----------------------------------------------------------------
\section{Introduction}

Electrical energy markets play a crucial role in modern economies. Reliable and cheap energy supply is believed to be essential for the daily lives of citizens and business operations. With the evolution and changing needs of economies, a large number of developed countries abandoned monopolistic and government-controlled power systems in favor of decentralized market structures. Today, trade is conducted in multiple ways: through bilateral contracts, short-term spot markets organized in the form of energy exchanges, and futures markets. Spot markets typically take the form of day-ahead (DA) markets in which offers are placed for 24 hours of the next day. In many countries, they are complemented by an intraday market that allows trade up to a few minutes prior to delivery and helps to dynamically balance supply and demand. DA prices serve as a reference price for other types of contracts and therefore have a huge impact on the whole market.

Together with changes of the market structure, one could also observe dynamic development of new electricity generation technologies, in particular those utilizing renewable energy sources (RES). As RES generation is intermittent and depends on changing weather conditions, it introduces a lot of uncertainty into the trade. Varying generation combined with the constant need to balance demand and supply, and limited storage opportunities, leads to high volatility of electricity prices. Their average level changes according to the weather and demand. In unfavorable circumstances, they also exhibit spiky behavior, with both positive and negative jumps.

The growing complexity of electricity markets and the increasing exposure of market participants to various trading risks raise the need for reliable forecasts of both electricity prices and market fundamentals, such as the demand level or RES generation. When reviewing the literature on electricity price forecasting (EPF), it is evident that the primary focus is often on point forecasting, i.e. forecasting the expected value of prices \citep[][]{wer:06}. Various methods have been proposed and examined in that context, starting with simple linear regression and autoregression types of models \citep{wer:14}, through nonlinear models \citep[see e.g.][]{wang_robust_2019,mac:uni:ser:20}, different estimation methods \citep{zie:16:TPWRS} up to artificial intelligence (AI) approaches \citep{lag:mar:des:wer:21}.  

In contrast to point predictions, probabilistic forecasts assess the entire distribution, thus allowing for assessing both the variable's level and the prediction uncertainty. 
Although more difficult to calculate, probabilistic predictions have gained popularity in the EPF literature \citep[][]{nowotarski_recent_2018, pet:etal:22}. The available methods can be classified into three main categories based on the way they express the uncertainty of the forecast: quantile prediction, density estimations, and ensemble predictions. As stated in \cite{gne:etal:08}, the distinction between different types of forecast is to some extent artificial. For instance, quantiles can be derived from an ensemble, and an ensemble can be generated using a fine approximation of the distribution. 
The literature discusses various methods for constructing probabilistic forecasts. Popular approaches are based on the analysis of forecast errors that come from a model used for point forecasting \citep[][]{kath_conformal_2021, mac:22} or direct modeling of a density of data with quantile regression \citep[QR, ][]{koe:05}.  It should be mentioned that to obtain a fine approximation of a continuous distribution, a substantial number of quantiles needs to be modeled, which makes an application of QR computationally burdensome. Moreover, the concept of quantiles describes primarily a univariate distribution and does not have a direct and intuitive extension to a multiple output case \citep{bre:cha:88}. Despite these obstacles, a very good QR prediction performance has been verified by a number of experiments, not
only in the EPF area \citep{kath_conformal_2021, uni:wer:21, mac:now:16, gai:gou:ned:16}. 

Whether in the context of point or probabilistic forecasting, typical techniques are designed to predict a single random variable. However, in many applications, the dependencies between events are important and should be carefully modeled \citep{gne:etal:08, pin:13}. For example, a wind farm manager can explore a correlation between wind generation and electricity price forecast errors to increase income \citep{lee_bivariate_2018}. An unexpected rise of RES generation is known to lead to a fall of DA and ID prices, which should be reflected in the offer curve.  Toubeau et al. \cite{toubeau_forecastdriven_2022} analyze a performance of a \textit{virtual power plant} (VPP) and shows how a multidimensional forecast of non-shiftable load, renewable generation, and electricity prices can be used in the management of the VPP. Finally, joint forecasting of electricity prices can be employed for designing a trading strategy that allows one to place offers on multiple electricity markets, similar to \cite{kumbartzky_optimal_2017}, \cite{mac:22}, and \cite{jan:woj:22}.

Therefore, in this research paper, instead of forecasting electricity prices and fundamental variables separately, we consider multidimensional probabilistic forecasting. There are two popular approaches that enable joint predicting of a set of time series. The first aggregates all endogenous variables into a vector and uses a single model to describe its behavior. A well-known example of such multidimensional models is \textit{vector autoregression} \citep[VAR,][]{lue:05}.  The approach is used in \cite{mac:22}, where behavior of electricity prices, total load, and RES generation is modeled with the structural VAR method. An alternative approach combines distributions coming from univariate models via, for example, copulas \citep[see][]{dur:15, gne:kat:14}. In such a case, the modeling procedure is divided into two steps. Firstly, time-series or AI models are applied to the individual variables. In the second step, the dependence between fitted errors is described by the copula. In the EPF literature, this approach has been successfully applied by \cite{ dur:gia:etal:22,toubeau_forecastdriven_2022, man:etal:19,ign:tru:16}.

In this study, we propose an alternative \textit{multiple split} procedure that is based on resampling methods. Similarly to \cite{ dur:gia:etal:22}, we use univariate models to predict the expected values of individual variables and a multidimensional distribution of forecast errors to calculate probabilistic forecasts. The method can be particularly useful for practitioners, as it may be combined with any unidimensional point-forecasting scheme. In this article, we explore time series models. However, it could also be merged with other structural or AI based techniques. 

The proposed method combines previous work of Lei et al. \cite{lei:etal:18} and Barber et al. \cite{bar:etal:21} and integrates \textit{multiple split conformal} predictions with a \textit{jackknife+} approach. Several names have been assigned to approaches that are based on the division of the sample into disjoint subsets: split method, leave-$k$-out (LKO), $d$ delete jackknife, or cross-validation (CV) \citep{efr:gon:83, vovk:etal:18, bar:etal:21}.
In the split approach \citep{lei:etal:18, kat:zie:18}, the sample is randomly divided into disjoint sets. The subsets resulting from the split are next used to estimate the parameters and calculate the forecast errors.  In this article, following \cite{bar:etal:21}, the outcome of each split is the probabilistic forecast of the variable of interest not just the distribution of errors. Hence, the method captures both the uncertainty that arises from estimation of parameters and the stochastic nature of the data. The \textit{multiple split}  described in this work enhances the existing literature in various directions:
\begin{itemize}
    \item[(i)] It extends the analysis from a univariate to a multivariate framework. The multidimensional property is particularly important when complex decision problems are considered that require approximation of a distribution of a function of random variables, e.g. their linear combination. In this case, the approach allows researchers to account for various sources of uncertainty that may cross-depend on each other.
    \item[(ii)] Unlike in previous articles such as \cite{ kat:zie:18}, random splitting is repeated multiple times to decrease the variability of the outcomes and reduce the dependence of the results on a particular division of the data. However, contrary to jackknife+ approach of Barber et al. \cite{bar:etal:21}, we do not consider all possible divisions of the original sample and hence reduce the computation time.
    \item[(iii)] The results obtained through the splits are stored as an ensemble rather then the distributions (parametric or a set of quantiles) as in \cite{lei:etal:18} or \cite{kat:zie:18}. Therefore, the approach does not require aggregation of outcomes obtained from individual splits with probabilistic forecast combination methods such as Bonferroni averaging \citep{lei:etal:18}. Moreover, the ensemble forecasts can be easily used to generate a prediction of any linear or non-linear function of the original data. In such a case, a new ensemble is constructed by applying the function to the set of multidimensional predictions.
\end{itemize}

The performance of the proposed method is evaluated with a dataset describing a German electricity market. First, it is applied for forecasting individual time series: day-ahead (DA) and intraday (ID) prices, total load, and RES generations. Next, it is used to predict the distribution of their function: the price spread and residual load (computed as the difference between load and RES). The accuracy of the forecast is assessed using various measures: the coverage of the prediction intervals, the \textit{continuous ranked probability score} (CRPS), and the reliability index. They show how well the probabilistic predictions resemble the out-of-sample distribution. Furthermore, given a correct calibration, CRPS assesses the sharpness of the distribution obtained. The results are compared to the well-established benchmarks: quantile regression and historical simulations. The second benchmark is extended to multidimensional forecasting to evaluate the potential gains from multiple splitting of the data. 
Finally, the method is used to support the decision process of moderate wind farms. The utility is assumed to decide about the quantity offered in the day-ahead market on the day preceding the delivery. The choice is made under uncertainty because at the time it is taken, the generator does not know neither its production level nor electricity prices. We show that joint forecasting of both market prices and fundamentals can be used to predict the distribution of future profit, and hence helps to design a trading strategy that balances the level of income and the risk. 

This article is structured as follows. Section \ref{sec:data} presents the main characteristics of the dataset explored in the study. Next, the forecasting methods are presented in Section \ref{sec:methods}. Section \ref{sec:simulation} describes the trading strategy of a small wind farm. Finally, the results of the empirical study are presented in Section \ref{sec:results}. Section \ref{sec:conclusions} provides conclusions.

%-----------------------------------------------------------------
\section{Data}\label{sec:data}
In this article, we use a dataset describing the EPEX SPOT market in Germany. The data span four calendar years, between October 1, 2015 and September 30, 2019, with an hourly resolution. They are roughly equally divided into training and test periods, with the first 728 days reserved for the initial calibration window and the remaining observations used for evaluation of the forecasts. The variables include day-ahead market prices (DA) and ID-3 intraday prices (ID), chosen as representative indicator of the average level of energy prices during continuous intraday trading; as well as actual and forecasted values of fundamental variables and, finally, closing prices of fuel futures contracts. The detailed description of all data used with units, sources, and notation is presented in Table \ref{tab:data_source}.

\begin{table}
\caption{Data sources and units.}
\label{tab:data_source}
\centering
\begin{tabular}{l|c|c|l}
\toprule
%\hline
Data & Notation & Units & Source\\
\hline
Day-ahead prices & $DA$ & EUR/MWh & \url{http://www.epexspot.com}\\
Intraday prices & $ID$ & EUR/MWh & \url{http://www.epexspot.com} \\
Load & $L$ & GWh & \url{https://transparency.entsoe.eu}\\
Wind generation & $W$ & GWh & \url{https://transparency.entsoe.eu}\\
PV generation & $S$ & GWh & \url{https://transparency.entsoe.eu}\\
Forecasted load & $FL$ & GWh & \url{https://transparency.entsoe.eu}\\
Forecasted wind generation & $FW$ & GWh & \url{https://transparency.entsoe.eu}\\
Forecasted PV generation & $FS$ & GWh & \url{https://transparency.entsoe.eu}\\
API2 Coal futures price & $C$ & EUR & \url{finance.yahoo.com}\\
TTF Gas futures price & $G$ & EUR & \url{www.eex.com}\\
%\hline

\bottomrule
\end{tabular}
\end{table}

\begin{figure}\label{fig:data_prices}
    \centering
    \includegraphics[width=1.1\textwidth]{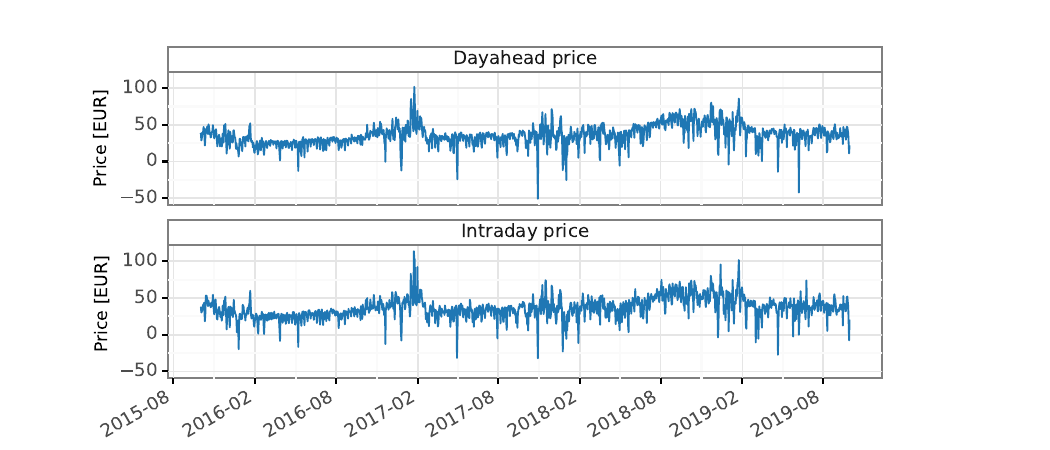}

    \caption{Germany, plots of market prices.}
\end{figure}

\begin{figure}\label{fig:data_res}
    \centering
    \includegraphics[width=1.1\textwidth]{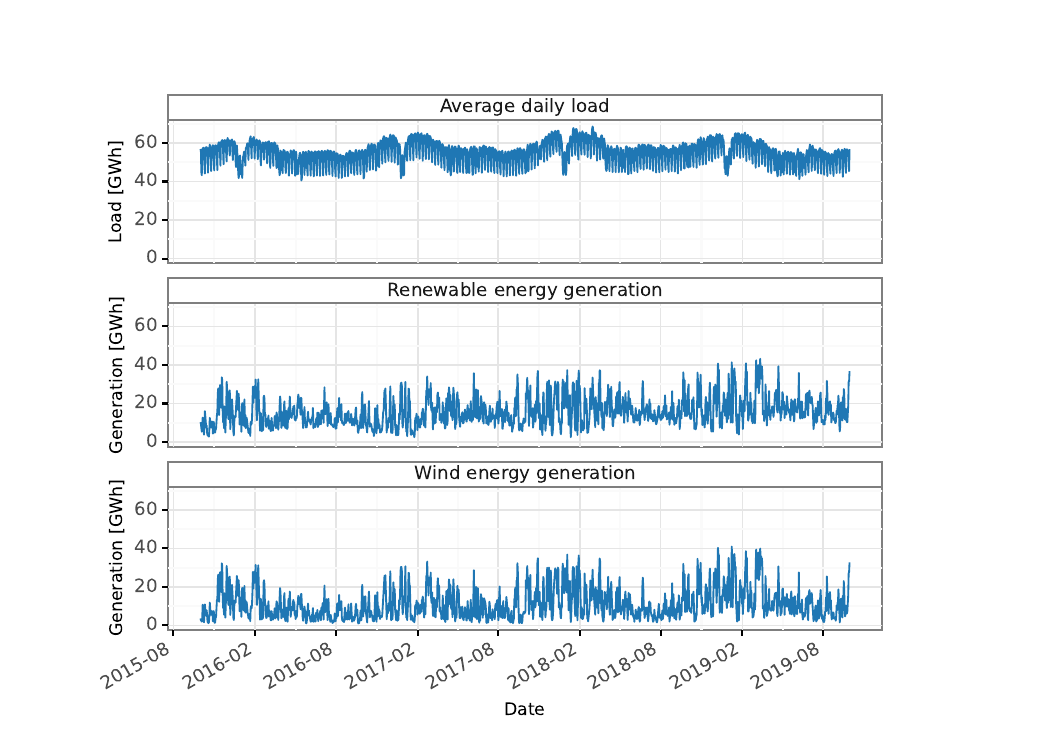}

    \caption{Germany, plots of total load and renewable energy generation.}
\end{figure}

The average daily values of the variables most relevant for the analysis, which are the prices in both the short-term markets and the generation structure, are plotted in Fig. \ref{fig:data_prices} and Fig. \ref{fig:data_res}, respectively. The price time series exhibits multiple features that differentiate the electrical energy market from most other commodity markets. They follow cyclical fluctuations that result from seasonal patterns of demand and yearly changes in weather. When prices on short-term markets are considered, it can be observed that they are typically highly volatile and strongly correlated. The average values across markets are similar. For the period analyzed, they are equal to 36.11 EUR/MWh for the day-ahead market and 36.24 EUR/MWh for the intraday market. However, the ID market is characterized by a higher variance, with a standard deviation of 18.31 EUR/MWh compared to 16.93 EUR/MWh on the DA market.  Finally, unlike in the case of other commodities, electricity prices are subjected to spikes. Although in most days the average price stays between 10-30 EUR/MWh, sudden jumps occur when the price increases over 50 EUR/MWh or falls below zero. 

Figure \ref{fig:data_res} complements the information on the short-term electricity market and presents the load level (\textit{top panel}, Fig. \ref{fig:data_res}), the RES generation (\textit{middle panel}, Fig. \ref{fig:data_res}) and the wind generation (\textit{bottom panel}, Fig. \ref{fig:data_res}). It can be observed that the load behavior is regular and exhibits strong weekly and yearly seasonality. 
In the case of RES generation, the level of production varies mainly due to short-term intermittency and changing weather conditions. These fluctuations have a significant effect on the wholesale market and lead to an increase of the price variability. The impact is the stronger, the larger becomes the share of RES in the generation mix. Within the considered time period, the installed wind energy capacity in Germany increased by approximately 36\%, from 44.58 GW in 2015 to 60.75 GW in 2019\footnote{Source: energy-charts.info}. The actual average monthly generation varied between approximately 10\% during summers (June to August) and 40\% during winters (December to February). The share of RES in the generation mix increased even more over the period due to investment in solar generation. In 2019, RES has been estimated to account for 44.8\% of the total energy generation in Germany.

All time series were pre-processed to account for time zone changes, as in \cite{wer:06}. The missing values (corresponding to the transition from winter to summer) were replaced by the arithmetic averages of the two nearest values. The doubled values (corresponding to the change from summer to winter) are replaced by their arithmetic mean.

%-----------------------------------------------------------------
\section{Forecasting methods}\label{sec:methods}

In this paper,  electricity prices and variables that describe the generation structure in consecutive hours are interpreted as separate time series (products). Their point forecasts are calculated with univariate models. Although the structure of the models remains unchanged throughout the day, the values of the parameters are estimated independently for each hour. Finally, in order to resemble the true trading problem, it is assumed that all the computations are performed in the morning at 11:00, hence only the information available at this time is used.

%-----------------------------------------------------------------
\subsection{ARX models}\label{sec:arx_models}

Autoregressive models with exogenous variables (ARX) are commonly used in the EPF literature \citep{nowotarski_recent_2018}. In this type of model, an expected value of an endogenous variable is described as a linear function of its past realizations (an autoregressive component) and a set of explanatory variables. When the structure of the model is predetermined and is not subjected to statistical selection and verification, it can be viewed as an expert model.  Here, we adopt specifications used for forecasting of electricity prices, RES generation, and the total load described by Maciejowska et al. \cite{maciejowska_enhancing_2021}. Additionally, two models are proposed that describe a residual load, defined as the difference between total load and the renewable energy generation ($RL_{t,h} = L_{t,h} - RES_{t,h}$), and a price spread ($SP_{t,h}=DA_{t,h}-ID_{t,h}$). 

Let us first consider models that explain the generation structure: total load, RES and wind generation, and residual load. Following \cite{maciejowska_enhancing_2021}, the load level is described by an equation below: 

\begin{equation}
\label{eq:ARX_L}
\begin{split}
    L_{t,h} = &\alpha_h  +\underbrace{\theta_{h,1} L^*_{t-1,h}+\sum_{p \in \{2,7\}} \theta_{h,p} L_{t-p,h}}_{\text{AR component}}+ \underbrace{\beta_{h,1} FL_{t,h} + \beta_{h,2} FRES_{t,h}}_{\text {Forecasts of fundamentals}}\\
    & + \underbrace{\beta_{h,3} FL_{t,ave}+ \beta_{h,4} FL_{t,max}+\beta_{h,5} FL_{t,min}}_{\text{Daily statistics}} +  \varepsilon_{t,h},
\end{split}
\end{equation}

The model consists of three main components: an autoregressive component with lags $p \in \lbrace 1, 2, 7 \rbrace$ that captures short-term dependencies and weekly seasonality, TSO forecasts of fundamental variables ( $FL_{t,h}$ and $FRES_{t,h} = FW_{t,h} + FS_{t,h}$), as well as the average, minimum, and maximum forecasted load within day $t$.
Because forecasts are performed in the morning at 11:00, information on the load level for hours later than 10:00 is not available. In this case, $L_{t-1,h}$ is replaced by its TSO forecast. Therefore, in the regression (\ref{eq:ARX_L}), a variable $L_{t-1,h}$ is replaced by $L^*_{t-1,h}$, which is constructed as follows:

\begin{equation}
\label{eq:L*}
L^*_{t,h}  = \left\lbrace
	\begin{aligned}
    & L_{t,h}\quad \text{ if}\ h\leq 10, \\
    & FL_{t,h}\quad \text{if}\ h>10. 
    \end{aligned}
    \right.
\end{equation}

When RES and wind generation are considered, the model structure is slightly different. First,  the AR component is shortened and contains only one lag, $p=1$. This is motivated by the fact that neither wind nor solar generation exhibits weekly seasonality. TSO predictions comprise not only the forecasts for hour $h$, but also for the previous and next hour, respectively. It can be noticed that for hours $h=1$ and $24$  some of the information is not available. In such a case, the corresponding variables: $FRES_{t,h-1}$, $FW_{t,h-1}$ or $FRES_{t,h+1}$, $FW_{t,h+1}$, are not included in the regressions. The models take the following form:
\begin{equation}
\label{eq:ARX_W}
    W_{t,h} = \alpha_h + \theta_{h} W^*_{t-1,h}+ \beta_{h,1} FW_{t,h-1} +\beta_{h,2} FW_{t,h} +\beta_{h,3} FW_{t,h+1} + \varepsilon_{t,h}, 
\end{equation}
\begin{equation}
\label{eq:ARX_RES}
    RES_{t,h} = \alpha_h + \theta_{h} RES^*_{t-1,h}+ \beta_{h,1} FRES_{t,h-1} +\beta_{h,2} FRES_{t,h} +\beta_{h,3} FRES_{t,h+1} + \varepsilon_{t,h}. 
\end{equation}
In order to prevent data leakage, for hours $h>10$ the renewable energy generation is replaced by its forecast:
\begin{equation*}
\label{eq:W*}
W^*_{t,h}  = \left\lbrace
	\begin{aligned}
    & W_{t,h}\quad \text{ if}\ h\leq 10, \\
    & FW_{t,h}\quad \text{if}\ h>10,
    \end{aligned}
    \right.
\end{equation*}
\begin{equation*}
\label{eq:RES*}
RES^*_{t,h}  = \left\lbrace
	\begin{aligned}
    & RES_{t,h}\quad \text{ if}\ h\leq 10, \\
    & FRES_{t,h}\quad \text{if}\ h>10.
    \end{aligned}
    \right.
\end{equation*}

A model describing the residual load comprises the above model specifications. Similarly to the model (\ref{eq:ARX_L}), it contains three lags of load level, the predicted load, together with information on the daily statistics of the forecasted  load within the day $t$.  Like in the model (\ref{eq:ARX_RES}), it includes the information on the RES level on the previous day and the predicted RES generation in three consecutive hours: $h-1,h,h+1$, whenever available.

\begin{equation}
\label{eq:ARX_RL}
\begin{split}
    RL_{t,h} = & \alpha_h  +\theta^L_{h,1} L^*_{t-1,h}+\sum_{p \in \{2,7\}} \theta^L_{h,p} L_{t-p,h}  + \beta^L_{h,1} FL_{t,h}+ \beta^L_{h,2} FL_{t,ave}+ \beta^L_{h,3} FL_{t,max}\\
    & +\beta^L_{h,4} FL_{t,min}+\theta^R_{h,1} RES^*_{t-1,h}+ \sum_{i=-1}^1\beta^R_{h,2+i} FRES_{t,h+i} +  \varepsilon_{t,h},
\end{split}
\end{equation}

Finally, the last three models describe the behavior of market prices: DA, ID and their difference $SP$. These regressions are closely related to each other. They include seven deterministic dummy variables $D_t$, which capture weekly seasonality, an autoregressive component with lags $p \in \lbrace 1, 2,..., 7 \rbrace$, TSO forecasts of fundamental variables ($FL_{t,h}$ and $FRES_{t,h}$), as well as the average, minimum and maximum levels of $DA$ prices from the previous day, $t-1$. Furthermore, the model takes into account generation costs and includes fuel prices: coal ($C_{t}$) and gas ($G_t$).

\begin{equation}
\label{eq:ARX_DA}
\begin{split}
DA_{t,h}  & =\alpha_hD_{t}+\underbrace{\sum_{p \in \{1,\ldots,7\}} \theta_{h,p} DA_{t-p,h}}_{\text{AR component}}+\underbrace{\beta_{h,1}DA_{t-1,ave} +\beta_{h,2}DA_{t-1,min} +\beta_{h,3}DA_{t-1,max}}_{\text{Daily quantities}}\\
& +\underbrace{\beta_{h,4} FL_{t,h} + \beta_{h,5} FRES_{t,h}}_{\text {Forecasts of fundamentals}}+ \underbrace{\beta_{h,6} C_{t-1} + \beta_{h,7} G_{t-1}}_{\text {Fuel prices}} +  \varepsilon_{t,h}.
\end{split}
\end{equation}

\begin{equation}
\label{eq:ARX_ID_1}
\begin{split}
ID_{t,h}  & =\alpha_hD_{t}+\theta_{h,1} ID^*_{t-1,h}+\sum_{p \in \{2,\ldots,7\}} \theta_{h,p} ID_{t-p,h}
+ \beta_{h,1}DA_{t-1,ave}+\beta_{h,2}DA_{t-1,min}  \\ 
 & +\beta_{h,3}DA_{t-1,max} + \beta_{h,4} FL_{t,h} + \beta_{h,5} FR_{t,h} + \beta_{h,6} C_{t-1,h} + \beta_{h,7} G_{t-1,h} + \varepsilon_{t,h}.
\end{split}
\end{equation}
While the day-ahead prices for day $t$ are all known at the time of forecasting, the intraday prices are represented by a variable $ID^*$, defined analogously to (\ref{eq:L*}):

\begin{equation*}
\label{eq:ID*}
ID^*_{t,h}  = \left\lbrace
	\begin{aligned}
    & ID_{t,h}\quad \text{if}\ h\leq 10, \\
    & DA_{t,h}\quad \text{if}\ h>10 . 
    \end{aligned}
    \right.
\end{equation*}

The model of price spread has an analogous structure to eq. (\ref{eq:ARX_ID_1}). The only difference comes from the AR component, which includes lagged values of price spreads.
\begin{equation}
\label{eq:ARX_SP}
\begin{split}
SP_{t,h}  & =\alpha_hD_{t}+\theta_{h,1} SP^*_{t-1,h}+\sum_{p \in \{2,\ldots,7\}} \theta_{h,p} SP_{t-p,h}
+ \beta_{h,1}DA_{t-1,ave}+\beta_{h,2}DA_{t-1,min}  \\ 
 & +\beta_{h,3}DA_{t-1,max} + \beta_{h,4} FL_{t,h} + \beta_{h,5} FR_{t,h} + \beta_{h,6} C_{t-1,h} + \beta_{h,7} G_{t-1,h} + \varepsilon_{t,h}.
\end{split}
\end{equation}
The intraday prices for hours after 10:00 are not known and therefore the $SP^*_{t,h}$ takes the following form
\begin{equation*}
\label{eq:SP*}
SP^*_{t,h}  = \left\lbrace
	\begin{aligned}
    & SP_{t,h}\quad \text{if}\ h\leq 10, \\
    & DA_{t,h}\quad \text{if}\ h>10 . 
    \end{aligned}
    \right.
\end{equation*}

The use of ARX models can be motivated by their very low computational complexity and high interpretability coupled with a well-established performance in the literature. While the model specifications presented in this section are used for further algorithm steps in this paper, they may in practice be replaced with any point forecasting method.

\subsection{Quantile regression}\label{ssec:qr}

Quantile regression (QR) is a modeling approach that allows linking a selected quantile of a distribution, $\tau$, of an endogenous variable $Y_t$ with a vector of exogenous variables $X_{t}$. Since quantiles are well defined only for univariate processes, QR requires separate modeling of the variables of interest: electricity prices, load, and RES.
Let us denote by $Q_{\tau}(Y_t)$ the $\tau$ quantile of $Y_t$. Then QR assumes that
\begin{equation}
	Q_{\tau}(Y_t) = X_t\theta_{\tau},
	\label{eq:QR}
\end{equation}
where $X_t$ is a $(1\times K)$ vector of explanatory variables and  $\theta_{\tau}$ is a $(K\times 1)$ vector of parameters. Notice that $\theta_{\tau}$ depends on $\tau$ and changes with the analyzed quantiles. To keep the results comparable with other presented approaches, the variables included in the model (\ref{eq:QR}) correspond to the ARX specifications from the previous Section \ref{sec:arx_models}.

The coefficients of (\ref{eq:QR}) can be estimated by minimizing the sum of \emph{pinball scores} ($PS$) throughout the calibration window. A $PS_t$ for a given period $t$ is defined as:
  \begin{equation}\label{eq:pinball}
    PS_{t}(\tau)= \begin{cases}
 (1-\tau)(Q_{\tau}(Y_t) - Y_{t})  & \text{ for } Y_{t} < Q_{\tau}(Y_t), \\
 \tau(Y_{t} - Q_{\tau}(Y_t)) & \text{ for } Y_{t} \geq Q_{\tau}(Y_t).
 \end{cases}
 \end{equation}

The estimation process can be carried out for 99 percentiles: $\tau=0.01,...,0.99$ and therefore QR can be used to approximate the entire distribution of $Y_{t}$. Here, QR is employed for the construction of prediction intervals (PI). A $PI$ with a nominal coverage $1-\alpha$ can be estimated as
\begin{equation*}
        PI_{1-\alpha}^{QR} = [\begin{array}{cc}
           Q_{\alpha/2}(Y_{T+1}),  & Q_{(1-\alpha/2)}(Y_{T+1}) \\
        \end{array} ],
\end{equation*}
where $Q_{\tau}(Y_{T+1})$ is a predicted quantile of $Y_t$ for the period $T+1$.

\subsection{Historical simulations}\label{ssec:historical}
\label{ssec:his}
Historical simulation is a direct method of constructing probabilistic forecasts that is widely studied in the literature \citep{kat:zie:18}. 
In order to obtain a multidimensional ensemble of forecasts, the method must be applied to all variables of interest at the same time. Let us denote by $Y_t\in R^{K}$ a $K$-dimensional vector of endogenous variables. In the case of electricity markets, the vector may include information about electricity prices and generation structure, for example $Y_{t} = [DA_{t,h}, ID_{t,h}, L_{t,h}, RES_{t,h}]$. Suppose that we want to calculate a forecast for a period $T+1$. Let us define a training set, $S_{train}$, as a window that covers periods preceding $T+1$, which is used to construct probabilistic predictions. The algorithm consists of the following steps:
\begin{enumerate}
    \item Calculation of point forecasts $\hat{Y}_{t}$  for $t\in S_{train}$ with a moving window approach. 
    \item Estimation of forecast errors: $e_{t}=Y_{t}-\hat{Y}_{t}$.
    \item Construction of the ensemble of predictions 
    $$\Psi=\{ y\in R^K: y=\hat{Y}_{T+1}+e_{t}\}$$
\end{enumerate}
In this research, point predictions used in steps 1 -- 3 are based on ARX models described in Section \ref{sec:arx_models}. It can be noticed here that although the variable $Y_t$ is a vector, its individual elements are described by different equations. However, thanks to a simultaneous computation of the predictions, the residuals $e_t$ maintain the correlation structure of the true forecast errors.

The collection of forecasts, $\Psi$, is next used to construct prediction intervals. In a classical form, a $PI$ with a nominal coverage $1-\alpha$ can be estimated as
\begin{equation}
        PI_{1-\alpha}^{hist} = [\begin{array}{cc}
           Q_{\alpha/2}(\Psi),  & Q_{(1-\alpha/2)}(\Psi)
        \end{array} ],
\end{equation}
where  $Q_{\tau}(\Psi)$ is the $\tau$  quantile of the pool $\Psi$.

%-----------------------------------------------------------------
\subsection{Multiple split method}\label{ssec:bootstrap}

In this research, we propose a \textit{multiple split} forecasting method. It is an extension of an approach known in the literature under the name  \textit{split conformal prediction} or \textit{inductive conformal inference} described by \cite{lei:etal:18, vovk:etal:18, bar:etal:21, kath_conformal_2021}. The main idea of this forecasting scheme is to use a random split of the data to construct probabilistic predictions. First, the training data is divided into two disjoint windows: estimation and calibration.
The first (estimation) subset is used to estimate the model parameters, which are then applied to calculate point predictions of the observations both within the training window and out-of-sample (i.e. target point prediction). The forecast errors are then estimated by computing the difference between the actual observations and their predictions in the calibration subset. Finally, the errors are used to approximate the distribution of the dependent variable. The way the are explored depends on the adopted methodology. When classical resampling methods are considered, they are used to directly approximate the quantiles of the distribution. In the context of conformal predictions, their absolute values are used to construct prediction intervals. Since our interest goes beyond prediction intervals, the first approach is adopted in this paper.

The \textit{multiple split} approach extends previous works in various directions. Firstly, the random split is conducted multiple times in order to improve the forecast accuracy and decrease the variability of the outcomes. However, unlike in \textit{leave-$k$-out} or \textit{delete-$d$ jackknife} approaches \citep[see][]{shao_general_1989}, we do not consider all possible divisions of the sample. This would significantly increase the computational complexity, but add little to the prediction quality. Second, forecast errors from the training window are used to directly construct an ensemble of forecasts rather than to estimate the distribution quantile. Therefore, the final ensemble is constructed by adding the results from individual splits, and there is no need for an intermediary step of averaging the quantiles or prediction intervals as in \citet{lei:etal:18}. Finally, the approach is applied to predict a multidimensional random variable. We believe that resampling methods are of particular use in this context. They do not require parametric modeling of the multidimensional distribution, and at the same time maintain the correlation structure of forecasts and forecast errors.

\begin{figure}[ht]
    \centering
    \begin{tikzpicture}[
% line crossing handling - white line preaction
on top/.style={preaction={draw=white,-,line width=#1}}, 
on top/.default=4pt,
% node definition
whitenode/.style={rectangle, draw=black, fill=green!5, minimum size=30mm, text centered},
graynode/.style={rectangle, draw=black, fill=black!10, minimum size=30mm, text centered},
stripednode/.style={rectangle, draw=black, preaction={fill, green!5}, pattern={north east lines}, pattern color=black!20, minimum size=30mm, text centered},
% allow for text breaking
every text node part/.style={align=center},
% define calibration window drawing as one picture
calwindowstripes/.pic={%
\draw[fill=green!5] (0, 0) rectangle (12, 0.7);
\filldraw[draw=black, fill=black!10] (1,0) rectangle (2,  0.7)
  (2.5,0) rectangle (3,  0.7)
  (4.5,0) rectangle (6,  0.7)
  (7,0) rectangle (7.5,  0.7)
  (9,0) rectangle (10.5,  0.7)
  (11,0) rectangle (11.5,  0.7)
;},
]

% nodes

\pic[local bounding box=calwindow, draw=black, 
minimum width=120mm, minimum height=7mm] (calwindow) {calwindowstripes};
\node[above] at (current bounding box.north) {Sample};
\node[draw=black, preaction={fill, green!5}, 
pattern={north east lines}, pattern color=black!20, 
minimum width=7mm, minimum height=7mm, right=of calwindow] (targday) {$T+1$};
\node[whitenode] (parameters) [below=of calwindow, xshift=-45mm]{parameter \\ estimates \\ $\hat{\theta}_i$};
\node[graynode] (calforecasts) [below right=of parameters] {point forecasts \\
\\
                                                            $\hat{Y}_{t,i}=X_{t}\hat{\theta}_i$ };%\\ 
                                                           % $\ldots$ \\ 
                                                           % $t\in S_{train,i}$};
\node[graynode] (calerrors) [right=of calforecasts] {forecast errors\\
\\
                                                    $e_{t,i} = Y_t -\hat{Y}_{t,i} $};
\node[stripednode] (targforecasts) [above right=of calerrors] {point forecasts \\ 
\\
                                                                $\hat{Y}_{T+1,i} =X_{T+1}\hat{\theta}_i $};
\node[stripednode] (derforecasts) [right=of calerrors] {ensemble \\ 
                                                        %forecasts \\ 
                                                        \\
                                                         $\hat{Y}_{T+1,i}+e_{t,i}$}; 
%                                                        derivative \\ 
%                                                        forecasts \\ 
%                                                       $f\left(\hat{X}_d,\ \hat{Y}_d\right)$};

% lines
%  calibration window
%   white arrows
\draw[thick, Circle-] (0.5,0.4) |- (2.25, -0.5); % right side
\draw[thick, Circle-Stealth] (2.25, 0.4) -- (2.25, -1); % center
\draw[thick, Circle-] (3.75, 0.4) |- (2.25, -0.5); % left side
%   gray arrows
\draw[thick, Circle-Stealth] (5.25, 0.4) -- (5.25, -5);
\draw[thick, Circle-] (7.25, 0.4) |- (5.25, -0.5); % left side
\draw[thick, Circle-] (9.75, 0.4) |- (7.25, -0.5); % left side

%   small arrow to d
\draw[thick, -Stealth] (calwindow) -- (targday);
% diagram part
\draw[thick, -Stealth] (parameters) |- (calforecasts);
\draw[thick, -Stealth] (calforecasts) -- (calerrors);
\draw[thick, -Stealth] (calerrors) -- (derforecasts);
% \draw[ultra thick, draw=white, -Stealth] (parameters) -- (targforecasts);
\draw[on top, thick, -Stealth] (parameters) -- (targforecasts);
\draw[thick, -Stealth] (targforecasts) -- (derforecasts);

\end{tikzpicture}
    \caption{Schematic illustration of the algorithm.}
    \label{fig:schematic}
\end{figure}
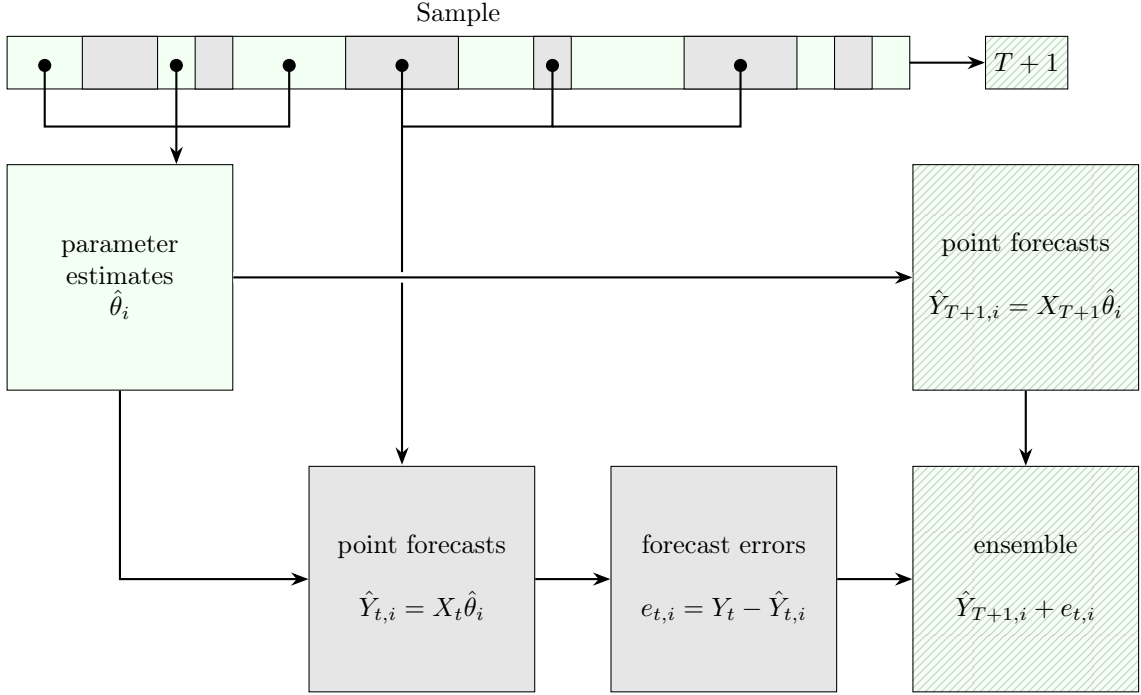

In the \textit{multiple split} method, similar to the historical approach, all variables are forecasted jointly and hence are collected in a $K$ dimensional vector $Y_t$. Suppose that we observe a sample $S$ that includes periods $t=1,...,T$ and we want to calculate a forecast for a period $T+1$. Each iteration of the proposed algorithm consists of $N$ independent splits. The procedure for a single split, $i=1,...,N$, is presented in Fig.\ref{fig:schematic}. On the graph, the estimation subset and associated steps are marked in green. Gray shading represents calibration periods. Striped boxes show the final forecasts. The placement of boxes indicates the order of operations (from left to right and from top to bottom), while the arrows show the dependency on the results from previous steps.
The $i$th split consists of the following steps:
\begin{enumerate}
    \item The sample is randomly divided into estimation ($S_{estim}^{(i)}$, green colour Fig. \ref{fig:schematic})  and calibration ($S_{calib}^{(i)}$, grey colour Fig. \ref{fig:schematic}) subsets such that $S=S_{estim}^{(i)}\cup S_{calib}^{(i)}$ and $S_{estim}^{(i)}\cap S_{calib}^{(i)} = \emptyset$.
    \item The data in the estimation window, $S_{estim}^{(i)}$, is used to estimate the parameters of models used for forecasting, $\hat{\theta}_i$, which are next employed to calculate predictions for periods $t\in \{S_{calib}^{(i)}, T+1\}$: $\hat{Y}_{t,i}=X_t\hat{\theta}_i$
    \item For each observation in the calibration window, $t\in S_{calib}^{(i)}$, the forecast error is calculated as $e_{t,i}=Y_t-\hat{Y}_{t,i}$
    \item The ensemble of predictions is constructed as 
    $$\Psi_i=\{ y\in R^K: y=\hat{Y}_{T+1,i}+e_{t,i}, t\in S_{calib}^{(i)}\}$$
   
\end{enumerate}
Steps 1 -- 4 are repeated $N$ times, and new sets of forecasts are added to the pool
\begin{equation}
    \Psi = \bigcup\limits_{i=1}^{N} \Psi_i.
\end{equation}
 Finally, analogously to the historical method, the ensemble of predictions is used to construct prediction intervals 
 \begin{equation}
        PI_{1-\alpha}^{MS} = [\begin{array}{cc}
           Q_{\alpha/2}(\Psi),  & Q_{(1-\alpha/2)}(\Psi)
        \end{array} ],
\end{equation}
 where $1-\alpha$ is the  nominal coverage level of PI.

As a result, the algorithm provides a set of forecasts that are derived from different splits of the data. In the proposed approach, we aggregate information rather than average prediction intervals as in \cite{lei:etal:18}. This method has several advantages. First, by aggregating the ensembles instead of the distributions, we do not need to decide on the averaging method for probabilistic (e.g. quantile) forecasts \citep[see][]{lic:g-c:win:13,ber:zie:21,nit:wer:23}. Second, the Bonferroni averaging method applied by Lei et al. \cite{lei:etal:18} to construct prediction intervals limits the number of splits, $N$, that can be used, as it requires estimation of a $\alpha/2N$ and $1-\alpha/2N$ quantiles of data. Finally, it allows us to use a wider variety of measures for assessing the forecast accuracy.

\subsection{Forecast evaluation} \label{ssec:forecast:eval}

Probabilistic forecasts discussed in previous sections represent two popular types of prediction: quantile forecasts and ensemble forecasts. Although it is possible to estimate quantiles from a set of predictions, it is more difficult to generate a diversified ensemble from a set of quantiles. Therefore, in this article, we use two separate approaches to assess the forecast accuracy that are dedicated to one of these types of predictions.

First, using QR, historical simulations, or the multiple split method, we approximate the distribution of the variables of interest with 99 quantiles and construct prediction intervals of nominal coverage 80\%, 90\%, 95\% and 98\%. The precision of PIs is evaluated with the \textit{PI coverage probability}  (PICP). The measure is based on the average number of observations in the testing window that fall into the PI. For an hour $h$, it is calculated as
\begin{equation}
    PICP_h = \frac{1}{T}\sum_t \mathbb{I}({Y_{t,h}\in PI_{t,h}}),
\end{equation}
where $\mathbb{I}(Y_{t,h}\in PI_{t,h})$ is an indicator function that takes value 1 when the variable $Y_{t,h}$ falls to the prediction interval $PI_{t,h}$ and zero otherwise. To obtain the final value, the $PICP$s for individual hours are averaged
\begin{equation}
    PICP = \frac{1}{24}\sum_h PICP_h.
\end{equation}
From the definition, this empirical coverage should be as close as possible to the nominal one. To evaluate whether $PICP_h$ is sufficiently close to the nominal coverage, we perform a Kupiec test \citep{kup:95} for each hour of the day separately. The null hypothesis says that the empirical coverage equals the nominal level, whereas under the alternative the $PICP_h$ differs significantly from $1-\alpha$. In this article, we report the percentages of hours for which we were unable to reject the null at the significance level 5\%. Therefore, the closer the measure is to one, the more successful the prediction method is in providing PIs of a predefined probability level. 

Finally, to assess the quality of the estimated 99 quantiles, we use CRPS described by Gneiting et al. \cite{gne:bal:raf:07} that evaluates both the calibration and the sharpness of the distribution. The measure is defined as an integral of the the pinball score defined in eq. \ref{eq:pinball} over the entire predictive distribution, and can be approximated for quantile forecasts as the arithmetic mean of pinball scores. For each of the 99 percentiles $\tau=0.01,...,0.99$, we calculate $PS_{t,h}(\tau)$.  The $CRPS$ of a given observation can then be calculated as
\begin{equation}
    CRPS_{t,h} = \frac{1}{99}\sum_{\tau} PS_{t,h}(\tau).
\end{equation}
The measure for the whole out-of-sample period is an average over all observations:
\begin{equation}
    CRPS = \frac{1}{T}\frac{1}{24}\sum_t\sum_h CRPS_{t,h}.
\end{equation}
The lower the value of $CRPS$, the better the approximation of the distribution. It should be noted here that the sharpness of the distribution has a significant impact on the measure. However, sharpness is a criterion that should be analyzed only when the calibration is correct. Furthermore, since the pinball score is a loss function used to estimate the QR, CRPS may favor the results that arise from this forecasting method. Thus, we believe that CRPS should be analyzed together with other measures, such as PICP.

In this article, we also use evaluation methods that are dedicated only to ensemble forecasts. In particular, we apply the reliability index \citep{gne:etal:08}, which assesses whether the ranks of actual observations within the pool of predictions have a distribution close to uniform.  

In the case of a univariate variable, $Y_{t,h}$, we denote by $r_{t,h}$ a proportion of ensemble forecasts smaller than or equal to $Y_{t,h}$. Since $r_{t,h}$ resembles a cumulative distribution function, it should have a uniform distribution. Let us divide an interval $[0,1]$ into $M$ equal bins: $B_1, ..., B_M$, and denote by $f_{j,h}$ the frequency of $r_{t,h}$ falling into bin $j$ in the out-of-sample period.  
\begin{equation*}
    f_{j,h} = \frac{1}{T}\sum_{t=1}^T\mathbb{I}(r_{t,h}\in B_j).
\end{equation*}
Then the reliability index can be calculated as follows
\begin{equation}\label{eq:RI:h}
    \Delta_h = \sum_{j=1}^M |f_{j,h} - \frac{1}{M}|.
\end{equation}
The lower the value of $\Delta_h$, the closer the empirical distribution of $r_{t,h}$ is to the uniform one. The final measure of the discrepancy is calculated as an average of $\Delta_h$ over 24 hours.
\begin{equation}\label{eq:RI}
    \Delta = \frac{1}{24}\sum_{h=1}^{24}\Delta_h
\end{equation}

To assess the accuracy of the multivariate probabilistic forecast, we use the idea of a multivariate rank histogram described by \citep{gne:etal:08}. Given the ensemble of forecasts, $\Psi$, and the verifying observation, $Y_0\in R^K$, the procedure consists of the following steps:
\begin{enumerate}
    \item First, the \textit{pre-ranks} are assigned, such that
    $$ \rho_j=\sum_{i=0}^M \mathbb{I}(Y_i\preceq Y_j),$$
    where $Y_i\in\Psi$ for $i=1,...,M$. Moreover,  $Y_i\preceq Y_j$ if and only if $Y_{i,k}\leq Y_{j,k}$ for all $k=1,...,K$.
    \item Next, \textit{the multivariate rank} is calculated as the rank of the pre-ranks. When two observations have the same pre-rank, the final rank is assigned randomly. Let us adopt the following notation:
$$ r_1 = \frac{1}{M}\sum_{j=0}^M \mathbb{I}(\rho_j<\rho_0),$$
$$r_2 = \frac{1}{M}\sum_{j=0}^M \mathbb{I}(\rho_j<\rho_0).$$
Then the final rank, $r$, is chosen randomly from the set $\{r_1+1,...,r_1+r_2\}$.
\item Similarly to the univariate case, to calculate the reliability index, divide an interval $[0,1]$ into $M$ equal bins and denote by $f_{j,h}$ the frequency of $r_{t,h}$ falling into bin $j$. Then the discrepancy indices can be calculated according to (\ref{eq:RI:h}) -  (\ref{eq:RI}).
\end{enumerate}

The multivariate rank histogram shows whether the multidimensional distribution is well calibrated to the data. It depends on the quality of the marginal distributions as well as on the ability to approximate the correlation between the variables. 

\section{Support of the decision process of a wind farm}\label{sec:simulation}

In this research, we show how the joint prediction of different market fundamentals can be used to support the decision process in the electricity market. We analyze the trading decisions of a company that owns wind farms spread throughout Germany. The amount of energy produced is assumed to be small enough not to influence market prices. Each day, wind turbines generate electricity, which is sold in the power energy exchange. The company has access to the day-ahead (DA) and intraday (ID) markets and does not speculate; i.e., it aims to sell the entire generated energy in either of these markets. Although purchasing energy on the intraday market may be necessary in the case of an overestimated generation forecast, this operation is never done intentionally.

The day before delivery, the company places bids on the DA market. The decision on how much to offer there has a significant impact on the trading outcome. Due to the intermittent nature of the wind, the company faces uncertainty about the level of generation that hampers the decision-making process.
If the company sells less electricity in the DA market than it produces, the remaining generation needs to be offered in the ID market. On the contrary, if it offers more than it generates, it needs to buy the missing production in the ID market. To construct the offer, the company may use predictions of both its generation and market conditions. Similarly to \cite{mac:22}, decision is described by a parameter $q \in [0, 1]$ that represents the fraction of the forecasted wind generation $\hat{W}$ offered in the day-ahead market. 

In this paper, it is assumed that the company faces some operational and maintenance costs $C_{O\&M}$, which influence the level of marginal profits. For simplicity, these costs are kept constant and expressed in EUR per 1~MWh of production. Other types of cost are interpreted as fixed costs and therefore do not influence the decision process. Then the profit of the company can be calculated as follows
\begin{equation}
\label{eq:profit}
\Pi(q)   =  \underbrace{q \hat{W}_{t,h} DA_{t,h}}_{\text{DA market}} + \underbrace{(W_{t,h} - q \hat{W}_{t,h}) ID_{t,h}}_{\text{ID market}} - \underbrace{W_{t,h}C_{O\&M}}_{\text{O\&M costs}}.
\end{equation}
Hence, the profit per 1MWh of generation is
\begin{equation}
\label{eq:average_profit}
\pi(q)   =  q \hat{w}_{t,h} DA_{t,h} + (1 - q \hat{w}_{t,h}) ID_{t,h}-C_{O\&M},
\end{equation}
where $\hat{w}_{t,h}$ shows how close is the predicted generation level to the actual one: $\hat{w}_{t,h} = \hat{W}_{t,h}/W_{t,h}$. When $W_{t,h}=0$ then the average profit is also set to zero.

It should be noted here that the level and distribution of income depend on three main sources of uncertainty associated with the unknown level of generation, $W_{t,h}$, and market prices: $DA_{t,h}$, $ID_{t,h}$. Since profit is a nonlinear function of these variables, its distribution is nontrivial. Resampling and simulation methods, such as the historical approach or the multiple split method, are of great help in estimating its probabilistic forecasts. The set of joint predictions of different market fundamentals can be used to construct an ensemble of future profits, which in turn can be used to approximate the income distribution.

\subsection{Bidding strategies}\label{ssec:strategies}

In this research, various trading strategies are considered. The benchmark  \textbf{naive strategy} assumes that the entire predicted generation is sold on the DA market. The profitability of this approach depends primarily on the accuracy of wind generation forecasts.  It is assumed that the company places an unlimited bid for the quantity $\hat{W}_{t,h}$ on the DA market and balances the position on the ID market.  Hence, the parameter $q$ is fixed and equal to $q=1$. This implies that the strategy responds only to the fluctuation of the generation but does not adjust to the market situation. 

Next, three data-driven strategies are proposed that explore probabilistic profit forecasts to design the optimal trading strategy. As mentioned above, the strategy is adjusted with a parameter $q$. For a grid of different values of $q$, the ensemble of predictions of $\pi(q)$ is constructed. The set of future profit values, $\Psi(q)$, is used to derive the optimization criteria and select the best value of $q$.

The first and most natural approach to choosing $q$ is to maximize the expected value of the profits, henceforth called the \textbf{expected profit strategy}. It is based on a point forecast of income, calculated here as the median of the ensemble $\Psi(q)$. The value of $q$ is chosen as the one for which the median is the highest. This strategy is expected to bring high income at the cost of increased risk.

The second \textbf{VaR strategy} is relevant for more risk-averse traders. It focuses on minimizing risk, commonly represented in business applications by Value-at-Risk (VaR). Here, we calculate the Value-at-Risk as the 5th percentile of the ensemble. Note that VaR is often expected to be negative and therefore presented as the absolute value of the quantiles of predicted or historical returns, in which case the quantity should be minimized. In this application, it is possible that VaR is positive (i.e., even the worst case scenario would not lead to losses), and therefore we do not apply absolute value, leading to higher VaR being more desirable.

The final strategy, called \textbf{Sharpe Ratio strategy}, aims to find a balance between maximizing profits and minimizing risk. To do this, the Sharpe ratio (expected profit divided by its standard deviation, \cite{sha:94}) is calculated for different values of $q$. The optimal $q$ is chosen so that the ratio is maximized. The Sharpe ratio strategy, contrary to two previous approaches, explores information about the entire predictive distribution instead of its selected quantile.

The three data-driven strategies can be summarized as follows:
\begin{enumerate}
    \item \textbf{Expected profit strategy} ($E\pi$): $q = \operatornamewithlimits{argmax}_{q \in [0, 1]} Q_{50\%}(\Psi(q))$,
    \item \textbf{VaR strategy} ($VaR$): $q = \operatornamewithlimits{argmax}_{q \in [0, 1]} \operatorname Q_{5\%}(\Psi(q))$,
    \item \textbf{Sharpe ratio strategy} ($SR$): $q = \operatornamewithlimits{argmax}_{q \in [0, 1]} \operatorname{SR}(q) = \operatornamewithlimits{argmax}_{q \in [0, 1]} \frac{\Bar{\pi}(q)}{\sigma_\pi(q)}$, where $\bar{\pi}(q)$ and $\sigma_{\pi}(q)$ denotes the average value and the standard deviation of the predicted profits in the ensemble $\Psi(q)$, respectively.
\end{enumerate}

It can be noted that in the above specifications, the value of the profit per 1 MWh of generation is used instead of the total profit. In addition, in the remaining parts of the article, the profit per 1 MWh of generation is used to evaluate the results of the experiment because $\pi$ does not depend on the scale of the company (as long as it does not have market power to impact electricity prices).

%-----------------------------------------------------------------
\subsection{Stopping rules}\label{ssec:stopping}

It is well documented in the literature and is visible in Fig. \ref{fig:data_prices} that electricity prices can fall below zero. The existence of very low or negative prices implies that even an optimal bidding strategy can cause losses. To avoid such a situation, we assume that the company is allowed to curtail the production. It means that it either stops the generation (the turbines are turned off) or stores electricity. The second solution becomes an attractive alternative as more and more investments are made in the development of energy storage systems. 

In our research, the production curtailment implies that no electricity is sold in any of the markets, which simulates turning off the turbines without storing the unsold generation. In the case of naive strategies, the trader may place a limited bid instead of an unlimited bid on the DA market and set a lower bound for a price at zero. We call this a \textbf{naive limited bid strategy}. 
In case of data driven strategies, more complex solutions are available that account for a risk aversion of company owners. We assume that the generation is stopped when a selected quantile of the profit distribution is less than zero.  It means that the company is engaged in the trade when  $Q_{\tau}(\Psi(q^*))\geq 0$  and curtails the production for $Q_{\tau}(\Psi(q^*))< 0$. The fraction of predicted generation offered on the DA market, $q^*$, is selected with one of the three approaches: $E\pi$, $VaR$  or $SP$  discussed in  the previous Section \ref{ssec:strategies} .

It can be noticed that a risk-neutral trader would likely base the decision on the center of the distribution (for example, a median), whereas a risk-averse trader would place bids depending on pessimistic scenarios (low quantiles). A trader who is reluctant to stop production may consider high quantiles that exceed 50\%. Finally, as the aversion to curtailment increases, the selected quantile may converge to one, which will represent the strategy that assumes daily trade (i.e. without stopping generation).

%-----------------------------------------------------------------
\subsection{Evaluation of trading strategies}\label{ssec:eval:trading}

The performance of the presented trading strategies is evaluated according to the level of average profit, the trading risk, and the frequency of generation curtailment. Hence, we provide the company with a broad perspective, which encompasses potential preferences and aversions of the trader.

When the income level is considered, two quantities are reported: the average profit and the average profit per trade. The main difference between these measures results from the approach towards days when the production is curtailed. The average profit includes the whole testing period and is calculated as the weighted mean of individual hourly profits:
\begin{equation}
    \bar{\pi} = \frac{1}{24T}\sum_{t=1}^T \sum_{h=1}^{24} \pi_{t,h}(q^*),
\end{equation}
where $q^*$ is an optimal value of the parameter selected with the analyzed strategy ($q^*$ changes over days and hours). On the contrary, the profit per trade, $\tilde{\pi}$, is calculated only for periods when the trade occurs, and therefore observations when the generation is curtailed are disregarded. These two measures, $\bar{\pi}$ and $\tilde{\pi}$, are different only for strategies with a stopping rule. 

To better understand the significance of the stopping rule, the frequency of trade is computed. It shows how often the stopping rule is not violated. We expect that a frequent curtailment of generation may result in a rise of the average profit per trade and a fall in the average total profit.

Finally, the trading risk is evaluated with the VaR measure calculated as the 5\% quantile of the average profit $\pi_{t,h}$. In the calculation of VaR only the days when the trade occurs are taken into account. Otherwise, VaR for many strategies with stopping rules will simply be zero, reflecting the profit on days when the generation is curtailed. However, in these periods, the company is not exposed to a trading risk.

%-----------------------------------------------------------------
\section{Results}\label{sec:results}

The results of the research are divided into two parts. First, Section \ref{ssec:accuracy} focuses on evaluating the statistical accuracy of probabilistic forecasts obtained with the methods discussed above. Next, in Section \ref{ssec:strategy}, the performance of the trading strategies based on the multiple split approach is assessed and the economic value of the predictions is measured. In the analysis, we adopt the following specifications:
\begin{itemize}
    \item The evaluation period consists of 730 observations.
    \item Two different window sizes are used to calibrate the parameters: 365 and 730 days, which correspond to one and two years of observations, respectively.
    \item In the multiple split approach, the sample is evenly split between the estimation and calibration parts. When the window of 365 days is considered, the calibration window consists of 183 observations.
    \item Two values of the number of splits are evaluated: $N=\lbrace1,20\rbrace$. The results of these specifications are denoted $MS(1)$ and $MS(20)$, respectively. 
\end{itemize}
The forecasting experiment is based on a rolling window scheme, in which each hour of the entire two-year validation period is predicted separately.

%-----------------------------------------------------------------
\subsection{Statistical accuracy of probabilistic forecasts} \label{ssec:accuracy}

To evaluate the accuracy of forecasts, we first consider the four fundamental variables that describe the electricity markets. DA price, ID price, total load, and RES generation. Next, the ability to predict a linear combination of these variables is evaluated: price spread (a difference between the DA and ID prices) and residual load (a difference between load and RES). In the case of QR models, the residual load regression is specified as eq. (\ref{eq:ARX_RL}). When the price spread is predicted, the exogenous variables included in the model are the same as in eq. (\ref{eq:ARX_DA}). For ensemble forecasts based on historical simulations or multiple split methods, the set of predictions is constructed as a function of individual elements (univariate forecasts) from the pool. Hence, there is no need to specify separate models for forecasting the linear combination of fundamental variables. 

%-----------------------------------------------------------------
\subsubsection{Forecasting of market fundamentals}

First, let us consider the probabilistic forecasts of the individual fundamental variables. The results presented in Table \ref{tab:pi} show the PI coverage probabilities for four  levels of PIs: 80\%, 90\%, 95\%, and 98\%, together with the average frequencies of the Kupiec test indicating a correct coverage and the levels of the CRPS measure. Forecasts are calculated using two different sizes of calibration windows: 365 and 730 days. The outcomes show that all the forecasting methods provide PIs that are too narrow and hence have PICP below the nominal level. However,  the PIs of the ensemble forecasts seem to be better calibrated to the data. In particular, for $T=365$, PICP of QR are much below the nominal coverage level, while for MS (20) they are close to $1-\alpha$.  This finding is confirmed by the results of the Kupiec test. In case of electricity prices (columns DA and ID, Table \ref{tab:pi}), the test indicates that the coverage level of the QR method is correct in less than 2\%  cases for T=365. At the same time, for MS(20) forecasting method, the PICPs are not statistically different from the nominal level in more than 90\%. When the longer calibration window is considered, the numbers are 43.33\% and 76.67\% for the ID prices, respectively. For load and RES, the differences between QR and MS are less substantial but still visible.   The proportion of cases for which the Kupiec test cannot reject the null is greater for MS (20) than QR by almost 63 and 20 percentage points for load and RES, respectively. 

When ensemble forecasts are analyzed,  it can be observed that historical simulations have PICPs closer to the nominal level than QR and MS(1) but worse than MS(20). Moreover, the results for different MS specifications show that an increase of the number of splits improves the calibration of the quantiles. There are also differences in behavior of the approaches for different sizes of calibration windows. MS methods work relatively better for shorter windows with $T=365$, while historical simulations have empirical coverage closer to the nominal level for $T=730$. In this respect, the QR performs similarly to the historical simulations.

Finally, the forecasting methods can be compared on the basis of CRPS. As the measure is based on the pinball score, which is minimized when estimating QR, it is not surprising that QR outperforms other approaches in this context. However, the differences between QR and MS(20) are moderate, and MS(20) has the lowest CRPS of all the methods for Load. Unlike in case of PICP, the historical simulations are outperformed by both QR and MS(20).  Also, the comparison of different window sizes leads to different conclusions than PICP showing that a shorter window is preferable for all the models.

\begin{table}[h]
    \caption{Accuracy measures of probabilistic forecasts based on quantile predictions: market fundamentals}
    \label{tab:pi}
    \centering
    \begin{tblr}{
  colspec={|c|cccc|cccc|},
  %hline{1,2,4},
  cell{1}{2,6} = {c=4}{c}, % multicolumn
  cell{3,10,17,24}{2} = {c=8}{c}, % multicolumn
}
    \hline
    Model & T=365 & & & & T=730 & & &  \\
    \hline
    Data & DA & ID  & Load & RES & DA & ID& Load & RES\\
    \hline
    PICP & QR & & & &  & & & \\
    \hline
     80\% &  71.66\% &  72.99\% &  76.39\% &  78.22\% &  75.52\% & 77.52\% & 76.04\% & 79.09\% \\
     90\% &  82.28\% &  83.54\% &  86.78\% &  88.28\% &  86.06\% & 87.10\% & 87.18\% & 89.45\%\\
     95\% &  87.71\% &  89.46\% &  92.49\% &  93.41\% &  91.41\% & 92.06\% & 93.23\% & 94.32\% \\
     98\% &  90.72\% &  92.06\% &  96.00\% &  96.72\% &  95.41\% & 96.11\% & 96.54\% & 97.29\%\\
    \hline
    Kupiec (\%)& 1.67\% &  0.00\% &  25.83\% &  74.17\% &  17.50 \% &  43.33\% &  35.83\% &  90.83\%   \\
    CRPS & 1.8692  &  2.5276 &  1.9691 &  1.8576 & 2.0067 & 2.6453 & 2.2387 & 1.9021\\
    \hline
    PICP & Historical & & & &  & & &  \\
    \hline
     80\% &  75.73\% & 76.28\%  & 77.32\% & 77.55\% &  75.47\% & 76.50\%  &  77.45\% & 79.02\%\\
     90\% &  86.31\% & 86.33\%  & 87.13\% & 87.50\% &  87.57\% & 88.11\%  &  88.56\% &  89.27\%\\
     95\% &  92.55\% & 92.80\%  & 93.09\% & 93.50\% &  93.78\% & 93.87\%  &  93.76\% &  94.70\%\\
     98\% &  96.68\% & 96.72\%  & 97.03\% & 97.35\% &  97.65\% & 97.48\% &  97.35\% &  97.72\%\\
    \hline
    Kupiec (\%)&  20.83\% & 33.33\% & 54.17\% & 74.17\% & 45.00\%  & 55.00\%  &  84.17\%   &  98.33\%\\
    CRPS &  2.0154& 2.7723 & 2.0188 & 1.9258  &  2.1130  & 2.7387  & 2.2521 & 1.9440 \\
        \hline
                PICP & MS(1) & & & &  & & &  \\
    \hline
     80\% & 73.10\% & 74.08\% & 77.00\% & 77.19\% & 72.60\% & 74.83\% & 75.05\% & 77.34\%\\
     90\% & 84.10\% & 85.36\% & 87.75\% & 87.42\% & 84.69\% & 85.51\% & 87.03\% & 88.07\%\\
     95\% & 91.36\% & 92.24\% & 94.07\% & 93.32\% & 91.51\% & 92.04\% & 93.15\% & 94.02\%\\
     98\% & 96.61\% & 96.92\% & 97.66\% & 97.15\% & 96.55\% & 96.86\% & 96.87\% & 97.41\%\\
    \hline
    Kupiec (\%)& 10.83\% & 13.33\% & 70.83\% & 64.17\% & 8.33\% & 20.00\% & 28.33\% & 73.33\% \\
    CRPS & 2.0749 & 2.7761 & 2.0129 & 1.9077 & 2.1627 & 2.7905 & 2.2774  & 1.9404\\
    
    \hline
        PICP & MS(20) & & & &  & & &  \\
    \hline
     80\% & 80.07\% &  80.76\%  &  78.83\% & 78.61\% &  76.99\% & 78.50\%  &  76.19\% & 78.12\%\\
     90\% & 90.13\% &  90.33\%  &  89.75\% & 88.53\% &  87.95\% & 88.34\%  &  88.14\% & 88.62\%\\
     95\% & 94.78\% &  94.95\%  &  95.03\% & 93.81\%&  93.25\% & 93.63\%  &  93.68\% & 94.25\%\\
     98\% & 97.67\% &  97.92\%  &  97.90\% & 97.33\% &  97.51\% &  97.49\% &  97.05\% & 97.58\%\\
    \hline
    Kupiec (\%)&  90.00\%  &  98.33\%  & 98.33\%   & 92.50\% &  60.00\%  &  76.67\%  & 51.67\%   & 81.67\% \\
    CRPS &  1.9090  & 2.5653  & 1.9689  & 1.8843 & 2.0780  & 2.6873  & 2.2542  & 1.9284\\
    
\hline

    \end{tblr}
\end{table}

Next, let us take a deeper look at the ensemble methods. Table \ref{tab:ri} presents the reliability index (RI)  for both univariate and multivariate analysis. It can be noted that the historical simulation can be interpreted as the multivariate approach because it provides residuals that maintain the correlation structure of the forecast errors.
When MS approaches are considered,  we show first the results for cases where MS is applied independently to all variables (called \textit{Uncorr}. in Table \ref{tab:ri}). Second, the outcomes of the simultaneous forecasting of the fundamentals are presented as MS \textit{Corr}. Joint modeling allows to approximate the correlation of the residuals. Similarly to the evaluation of quantile predictions, the reliability index is shown separately for short and long calibration window sizes.

The results of the marginal distributions show that the MS(20) approach provides the lowest index value for all variables, except for the DA, $T=730$. Moreover, it is demonstrated that from the perspective of an individual variable, there are no differences between joint and separate applications of the MS method.  Therefore, there are no gains of multidimensional modeling. Finally, similar to PICP, the reliability index indicates that MS(1) is inferior to MS(20) and the historical approach. When the forecasts of the multidimensional distributions are evaluated (last column of Table \ref{tab:ri}), the results confirm the superiority of the MS(20) method. It provides the lowest value among all specifications $RI=0.3027$. Furthermore, substantial differences could be observed between approaches that account for the correlation of forecast errors or not. For example, for $T=365$ and MS(20), the reliability index is 0.3267 for uncorrelated residuals and 0.3027 for correlated ones.  This indicates substantial gains of using multidimensional modeling in joint forecasting of market fundamentals.

\begin{table}[h]
    \caption{Reliability index: univariate and multivariate analysis: market fundamentals}
    \label{tab:ri}
    \centering
    \begin{tblr}{
  colspec={|c|c|cccc|c|},
  %hline{1,2,4},
  cell{2,8}{3} = {c=5}{c}, % multicolumn
  cell{1,2,3}{1} = {c=2}{c},
  cell{8,9}{1} = {c=2}{c},
    cell{4,6,10,12}{1} = {r=2}{m}, % multirow
}
    \hline
    & & DA & ID  & Load & RES & All \\
    \hline
    Model & & T = 365 & & & & \\
    \hline
     Historical&   &  0.3386  &  0.3460 &  0.3363  &  0.3400 & 0.3350 \\
     \hline
     MS(1)& Uncorr.  & 0.4053   & 0.3797   & 0.3415   &  0.3380  & 0.3660 \\
     & Corr.  &  0.4034 & 0.3769  & 0.3432  & 0.3393  & 0.3551 \\
     \hline
     MS(20)& Uncorr.  & 0.3315   &  0.3092  & 0.2923   &  0.2927  & 0.3267 \\
     & Corr.& 0.3375 & 0.3133  & 0.3049  & 0.2889   & 0.3027 \\  
\hline
     & & T = 730 & & & & \\
    \hline
     Historical&   & 0.3245  & 0.3185   &  0.3172  &  0.2904 &  0.3123\\
     \hline
     MS(1)& Uncorr.  &  0.3631  &  0.3401  & 0.3363   &  0.3129  & 0.3444 \\
     & Corr.  & 0.3598  & 0.3407  & 0.3308  & 0.3140  & 0.3327  \\
     \hline
     MS(20)& Uncorr.  &  0.3219  & 0.3112   & 0.3106   &  0.2941  & 0.3359 \\
     & Corr.& 0.3292 & 0.3112  & 0.3044  & 0.2893  &  0.3202 \\  
\hline
    \end{tblr}
\end{table}

%-----------------------------------------------------------------
\subsubsection{Forecasting of linear combinations of market fundamentals}

Since the analysis of the prediction accuracy of the fundamentals of the market indicates that MS(1) is inferior to MS(20), only the results for the latter are presented in the following sections. Next, to assess the gains from multidimensional forecasting, we analyze separately the outcomes for \text{MS} approach  computed separately (\textit{Uncorr.}) or jointly (\textit{Corr.}) for all four fundamental variables.  In case of the QR, the multidimensional modeling is not available, and therefore the price spread and the residual load need to be forecasted directly.

Table \ref{tab:pi:2} shows three measures that describe the accuracy of quantile predictions: PICP, proportions of the Kupiec test indicating a correct empirical coverage, and the average value of CRPS.  When the PICP is considered, the outcomes confirm that QR provides prediction intervals which are too narrow and therefore exhibit coverage much lower than the nominal levels. This property is particularly well visible when the rejection of the Kupiec test is analyzed. In the case of QR, the test does not allow one to reject the null of a correct coverage in only 4.15\% and 12.5\% of the cases for Price Spread and Residual Load, respectively ($T=365$). As the calibration window increases, the frequencies increase accordingly to 62.50\% and 24.17\%, but still remain much lower than for historical simulations and MS(20) with correlated errors. Similarly to previous outcomes, it can be observed that the accuracy of QR predictions increases as the sample size rises from 365 to 730 days. 

The empirical coverage measured by PICP of both ensemble methods is much closer to the nominal level than in the case of QR. For a shorter calibration window, the results of the Kupic test show that the coverage of the analyzed PIs is not statistically different from the nominal level in 73.33\% and 92.5\% cases for the MS (20) approach (MS,\textit{Corr.}). The frequencies are slightly lower for the historical simulation method and are approximately 20.00\% and 46.67\%. 

When the results of the CRPS measure are examined, the outcomes support previous findings, which show that the QR method provides a forecast with the lowest CRPS level. This indicates that the probabilistic forecasts of QR are sharper than those of other methods. 

Finally, as multidimensional models are compared with independent forecasting of fundamental variables, the results indicate the superiority of the first approach. The MS(20) method that does not account for the correlation of forecast errors (MS, \textit{Uncorr.}), provide too wide intervals with the empirical coverage far from the nominal level. The problem is particularly severe for the Price Spread, for which the PICP of 80\% PI exceeds 92\%. Furthermore, the performance of the method does not improve when a longer calibration window is used. In this case, the Kupiec measure falls below 6\% and 50\% for Price Spread and Residual Load, respectively.

\begin{table}[h]
    \caption{Accuracy measures of probabilistic forecasts based on quantile predictions: linear combination of market fundamentals}
    \label{tab:pi:2}
    \centering
        \begin{tblr}{
 colspec={|c|cc|cc|cc|cc|},
  %hline{1,2,4},
  cell{1}{2,4} = {c=2}{c}, % multicolumn
  cell{1}{6} = {c=4}{c}, % multicolumn
  cell{2}{6,8} = {c=2}{c}, % multicolumn
  cell{1}{1} = {r=3}{m}, % multirow
  cell{4,11}{2} = {c=8}{c}
}
    \hline
   & QR & &Historical & & MS(20) & & &\\
    \cline{6-9}
    & & & & &Corr. &  & Uncorr. & \\
    \hline
    & Spread & RL& Spread & RL& Spread &RL& Spread & RL\\
    \hline
    PICP& T=365 & & & & & & &\\
    \hline
     80\% &  74.71\%  &  75.88\% &75.83\% & 76.98\% & 82.61\% & 78.76\% & 92.74\% &  76.03\%\\
     90\% &  85.13\%  &  86.23\% &86.04\% & 86.92\% & 91.30\% & 89.16\% & 96.15\% &  87.01\%\\
     95\% &  90.67\%  &  91.82\% &92.00\% & 92.96\% & 95.55\% & 94.19\% &  97.87\% &  92.99\%\\
     98\% &  92.89\%  &  95.02\% &96.56\% & 97.12\% & 97.61\% & 97.22\% &  99.02\% &  97.15\%\\
    \hline
    Kupiec (\%) & 4.17\% & 12.50\% & 20.00\%& 46.67\% &  73.33\% & 92.50\% & 18.33\% & 65.00\% \\
        CRPS & 1.9176 & 1.9182  & 2.1840& 2.9201&  1.9870 & 2.8387 & 2.0984 & 3.2046\\
    \hline
    PICP& T=730 & & & & & & &\\
        \hline
     80\% & 78.85\%  & 75.74\% & 77.77\%& 76.86\% & 80.39\% & 76.90\% & 92.73\% & 76.59\%\\
     90\% & 88.23\%  & 86.82\% & 88.09\%& 87.08\%  & 89.90\% & 87.40\% & 96.48\% & 87.41\%\\
     95\% & 93.24\%  & 92.77\% & 93.92\%& 93.13\% & 94.55\% & 93.18\% & 98.04\% & 93.33\%\\
     98\% & 96.31\%  & 96.22\% & 97.35\%& 96.85\% & 97.61\% & 96.90\% & 99.16\% & 97.12\%\\
    \hline
    Kupiec (\%) & 62.50\% & 24.17\% & 72.50\%& 44.17\%&  91.67\% & 45.83\% & 5.83\% & 46.67\% \\
        CRPS & 1.9264 & 2.2044 & 2.2268& 3.2397&  1.9918 & 3.1097 & 2.1293 & 3.1101\\
    \hline
    \end{tblr}
\end{table}

\clearpage

%-----------------------------------------------------------------
\subsection{Economic value of forecasts}\label{ssec:strategy}

To assess the economic value of the newly proposed method, we use probabilistic predictions to support the decision process of a generation utility. As described in Section \ref{sec:simulation}, the company owns wind farms that are spread throughout Germany and needs to decide on the share of predicted production, $q$, offered on the DA market. The remaining part of the generation is sold on the ID market. In the case of wind generation,  operational and maintenance costs account for about 25\%-35\% of the \textit{levelized energy cost} (LCOE)  \citep{ren:etal:2021,nej:etal:2022}. It implies that O\&M costs vary between USD 10-30 per MWh \citep{cos:etal:2021}. In the article \cite{wis:bol:lan:2019}, the costs are estimated at USD 11 per MWh for onshore installations and are expected to decrease in the future. Since the exchange rate between USD and EUR has oscillated between 0.9-1 USD/EUR,  in this research we assume that $C_{O\&M}=10$.

\begin{figure}
    \centering
    \includegraphics[width=1\textwidth]{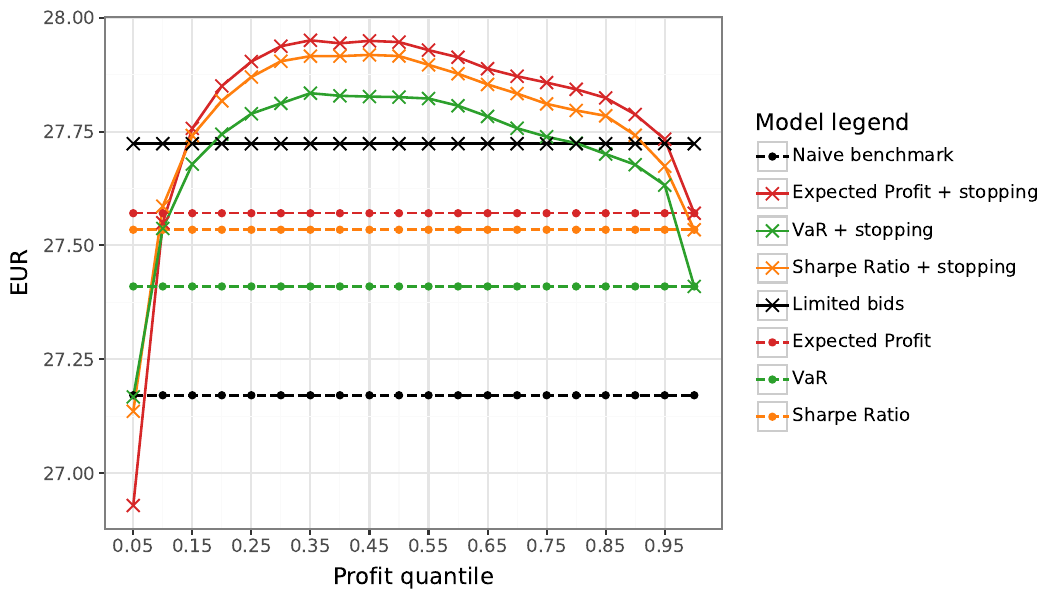}

    \caption{Average profit per 1~MWh of generation for two types of strategies: with production curtailment (marked with crosses) and without (marked with dots); depending on risk aversion level. Operation \& Management costs are 10 Euro/MWh.}
    \label{fig:income_total}
\end{figure}

First, let us analyze the average total income earned from different strategies. The results are presented in Fig. \ref{fig:income_total}, in which lines marked with dots show the outcomes of strategies without production curtailment, and the lines marked with crosses present the strategies that restrict trade. The naive benchmark, which assumes that all predicted generation is sold on the DA market, $q=1$, is depicted to facilitate comparison. It earns on average 27.27 EUR/MWh.  The results show that the adoption of data-driven strategies is profitable even when no curtailment is allowed. The highest revenue is achieved for the approach that aims to maximize the expected profit. A strategy optimizing the Sharpe ratio is only slightly worse, and brings 27.55 EUR/MWh instead of 27.56 EUR/MWh on average. When the results are compared with the benchmark, it can be seen that they bring 1.39\%  and 1.46\% more, respectively. 

The application of strategies that enable production curtailment requires the selection of a profit quantile that is used as a threshold in the stopping rule. Therefore, in the case of three data-driven approaches, revenue depends on the quantile $\tau$. For low values of $\tau$, generation is often reduced and, therefore, the company loses potential income. As the threshold quantile increases, the profit rises and exceeds both the naive benchmark and the limited bid. The strategies obtain their maximum of 27.95, 27.92, and 27.83 EUR/MWh for strategies based on $E\pi$, Sharpe ratio and VaR. Finally, as seen in Fig. \ref{fig:income_total}, income starts to decrease and falls to the level of the corresponding strategies without stopping the generation. The results indicate that the curtailment of production leads to a substantial increase of income. Compared to the naive benchmark, the limited bid strategy rises profits by 2.03\%. In the case of data-driven approaches, the increase is even greater and reaches 2.87\% for $E\pi$, 2.77\% for Sharpe ratio and 2.42\% for VaR strategies. This means that the stopping rule increases the profits by additional 2\%.

\begin{figure}
    \centering
    \includegraphics[width=1\textwidth]{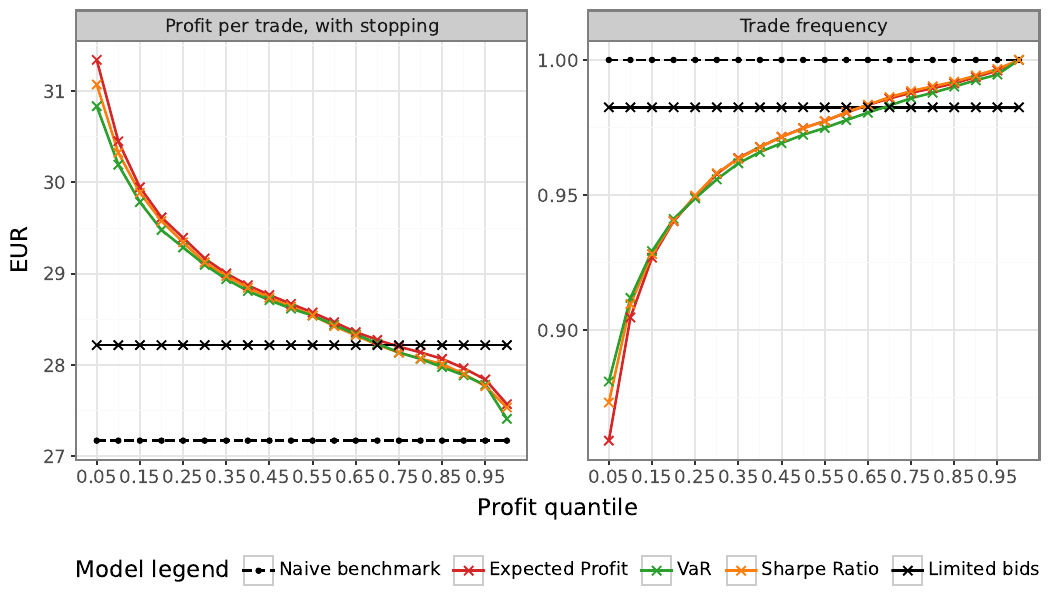}

    \caption{Performance of strategies with production curtailment: average profit per trade (left panel) and frequency of trade (right panel); Operation \& Management costs are 10 Euro/MWh.}
    \label{fig:income_trade}
\end{figure}

When the profit per trade is considered, it can be noticed that for strategies without curtailment, the average profit and the profit per trade are the same. Therefore, Fig. \ref{fig:income_trade} presents only the results of the strategies with stoppage of production. The profits of the naive benchmark are added to the plot for comparison. The results show that the risk averse strategies that are based on low quantiles of profits lead to an increase of the profit per trade to 31 EUR per MWh. This implies that it exceeds the income of the naive benchmark strategy by  15.34\%. As $\tau$ increases, the frequency of trade also rises, and the income per trade converges to the average total income presented in Fig. \ref{fig:income_total}. When the limited bid is considered, the profit per trade reaches 28.22 EUR/MWh and surpasses the naive benchmark by 3.85\%. However, income remains lower than for data-driven approaches for most thresholds, $\tau<70\%$. Only when the company becomes reluctant to curtailing the generation, the outcome of the limited offer exceed those of other strategies. Finally, it can be observed that among the data-driven approaches, the one that maximizes the expected value of profits is the most profitable. Similarly to previous results, it is followed by the Sharpe ratio strategy. The lowest income is earned by the VaR approach.

The performance of different strategies is summarized in Table \ref{tab:trade}, which shows the frequency of trade, average profits and average profits per trade relative to the naive benchmark.  The outcomes of approaches that allow for stoppage of productions depend on the quantile of profits used as the threshold. When $\tau$ is set equal to one (last column, Table \ref{tab:trade}), the strategy is equivalent to an approach without the stopping rule. The results indicate that data-driven methods lead to a reduction in the generation of 0.00\%--14.22\%. The greatest improvement is obtained for $\tau\in[0.3, 0.5]$, when the decrease in the generation frequency by less than 5\%  brings an increase in profits of more than 2.5\% in the case of average profits and 5-7\% in the case of average profit per trade. When the limited bid approach is considered, it can be seen that it leads to production curtailment in less than 2\% cases. At the same time, it brings an additional 2\% of the average profits. The strategies without stopping rule, shown in the last column of the table, confirm previous findings and demonstrate that they lead to a moderate increase in profits (between 0.88\% and 1.46\%).

\begin{figure}
    \centering
    \includegraphics[width=1\textwidth]{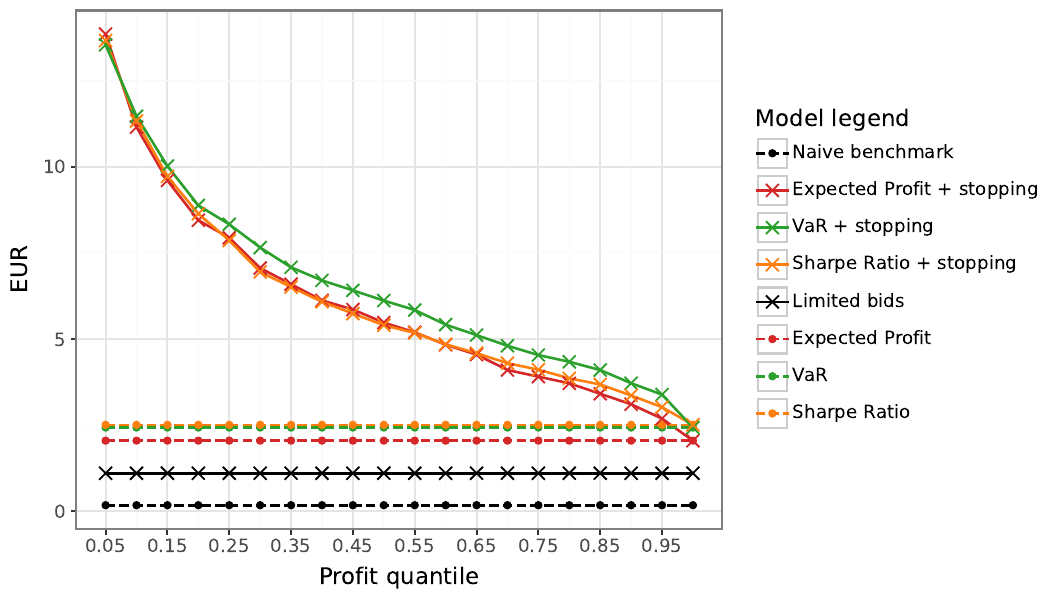}

    \caption{Value at Risk for two types of strategies: with production curtailment (marked with crosses) and without (marked with dots). Values depend on risk aversion level and are calculated only for the hours when the trade occurs. Operation \& Management costs are 10 Euro/MWh.}
    \label{fig:income_VaR}
\end{figure}

Economic evaluation is not complete without risk analysis. Here, 5\% VaR is used to measure profit in the case of a pessimistic scenario. The VaR for different strategies is presented in Fig.  \ref{fig:income_VaR}. The plot shows the results of the strategies without (dot markers) and with stopping rules (cross markers). First, it can be noticed that the naive benchmark is characterized with VaR slightly above zero. Other approaches guarantee a positive profit even at the bottom 5\% of the scenarios. When the income of strategies without stopping rule is considered, then the increase of VaR is moderate. Significantly lower risk is incurred by approaches that allow for generation curtailment. The limited bid approach provides VaR at 4.92 EUR/MWh. The results of data-driven strategies depend on the adopted threshold quantile, $\tau$, and vary between 2.05 and 13.88 EUR/MWh.
Finally, although the plot resembles the income per trade figure, it can be observed that the ordering of data-driven approaches is different. The least risky is the VaR strategy, while the $E\pi$ approach is characterized by the lowest value of VaR. Hence, more profitable strategies are at the same time more hazardous.

\begin{table}[h]
    \caption{Performance of trading strategies, relative to the naive benchmark}
    \label{tab:trade}
    \centering
        \begin{tblr}{
        colspec={|c|ccccc|c|},
        %hline{1,2,4},
        cell{1,3,7,8,12,13,17}{2} = {c=6}{c}, % multicolumn
        %cell{1}{2,5} = {c=3}{c}, % multicolumn
        %cell{8}{2} = {c=3}{c}, % multicolumn
        } 
    \hline

    Strategy& Quantile of profits, $\tau$& & & & & \\
    \cline{2-7}
    & 0.05& 0.30& 0.50& 0.70& 0.95& 1.00\\
    \hline
        &Frequency of trade & & & & &\\
    \hline
    $E\pi$ & 85.88\%& 95.75\%& 97.44\%& 98.61\%& 99.57\%& 100\% \\
    $VaR$& 88.05\%  & 95.64\%& 97.23\%& 98.30\%& 99.45\%& 100\%\\
    $SR$& 87.28\%   & 95.80\%& 97.43\%& 98.65\%& 99.64\%& 100\%\\
    \hline
    Limited bid &  0.9825& & & & & \\
    \hline 
    &Average profit   & & & & &\\
    \hline
    $E\pi$ & -0.95\%& 2.80\%& 2.87\%& 2.57\%& 2.11\%& 1.46\% \\
    $VaR$& -0.02\% & 2.37\%& 2.42\%& 2.22\%&  1.69\%& 0.88\%\\
    $SR$& -0.17\% & 2.71\%& 2.77\%& 2.03\%& 1.39\%& 1.39\% \\
    \hline
    Limited bid &  2.03\%& & & & &  \\
    \hline
        &Average profit per trade & & & & &\\
    \hline
    $E\pi$ & 15.34\%& 7.37\%& 5.58\%& 4.02\%& 2.55\%& 1.46\% \\
    $VaR$& 13.54\% & 7.03\%& 5.34\%& 3.98\%&  2.25\%& 0.88\%\\
    $SR$& 14.38\% & 7.20\%& 5.48\%& 3.87\%& 2.28\%& 1.39\% \\
    \hline
    Limited bid &  3.85\%& & & & &  \\
    \hline
    
    \end{tblr}
   
    Remark: Average profit and Average profit per trade are presented as the percentage difference between the selected strategy and the naive benchmark approach.
\end{table}

%-----------------------------------------------------------------
\subsubsection{Portfolio analysis}

\begin{figure}
    \centering
    \includegraphics[width=1\textwidth]{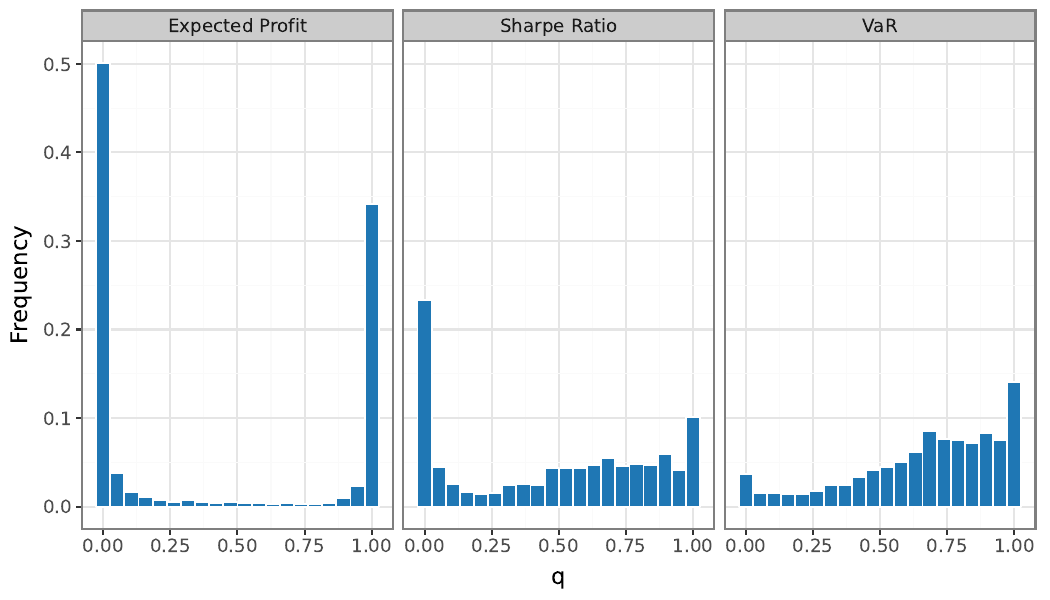}

    \caption{Histogram of optimal values of $q$ for different data-driven strategies: $E\pi$, $SR$ and $VaR$}
    \label{fig:histogram:q}
\end{figure}

The data-driven approaches presented above are characterized by different levels of profit and associated risks. Similarly to other commodities, contracts that bring greater income are associated with a higher level of uncertainty. Hence, the selection of the strategy depends on the risk appetite of the utility owners.
To make a proper decision, it could also be beneficial for the company to understand where the difference comes from. Fig. \ref{fig:histogram:q}  shows the histograms of the decision variable, $q*$ , for strategies without generation curtailment: $E\pi$ (\textit{left panel},  Fig. \ref{fig:histogram:q}), $SR$ (\textit{middle panel},  Fig. \ref{fig:histogram:q}) and $VaR$ (\textit{right panel},  Fig. \ref{fig:histogram:q}). It could be seen that they differ significantly in terms of the given recommendations. The strategy that maximizes expected profit selects in most cases extreme values of $q$. Hence, it offers the entire forecasted generation in the DA market or leaves the whole production for the ID market. Moreover, it chooses the ID market in 52\% of the cases, compared to only 36\% of the times when DA is selected.
In contrast, the $VaR$ strategy provides mainly diversified portfolios of contracts and recommends selling electricity in both markets. It selects $q=0$ or $q=1$ in 5\% and 17\% of the cases, respectively. Furthermore,  most of the time more than half of the predicted generation is offered on the DA market ($q\geq 0.5$). In 40\% of the periods, it recommends choosing $q\geq 0.8$.
As observed previously, the strategy $SR$ stays between $E\pi$ and $VaR$ . It selects intermediate values of $q$ in more than 60\% of cases. However, similar to the $E\pi$ approach, it is relatively often recommended to leave the entire generation for the ID market. It occurs in almost 26\% of the cases. 

In conclusion, data-driven selection of the market (setting $q=0$ or $q=1$) is profitable, but exposes the utility to a higher risk than choosing a balanced portfolio. Moreover, strategies that offer a greater proportion of forecasted generation in the DA market bring lower income, but at the same time reduce the trade uncertainty.

%-----------------------------------------------------------------
\section{Conclusions and discussion}\label{sec:conclusions}

In this article, we propose a \textit{multiple split} method to construct probabilistic forecasts of both one- and multidimensional random variables. The approach splits the training sample into two disjoint sets: estimation and calibration. The first subset is explored to estimate the model parameters, whereas the second is used to calculate the forecast errors.  Point forecasts based on calibrated parameters and prediction errors are used to construct an ensemble of forecasts. The splitting is repeated multiple times, and through merging the splits a final ensemble is constructed.

This work enhances the current body of literature on multidimensional probabilistic forecasting. The proposed approach combines and extends two methods: jackknife+ and split conformal predictions. In the MS forecasting scheme,  the random division of the sample is performed multiple times to improve the accuracy of the forecast and decrease the variability of the results. However, unlike other methods such as delete-$d$ jackknife, jackknife+ or leave-$k$-out , MS uses a small number of splits: here, the number varies between 1 and 20. This significantly decreases the computational complexity. Second, forecast errors from the training window are used to construct an ensemble of forecasts rather than to estimate the distribution quantiles. Hence, to obtain the final probabilistic forecasts, it is not necessary to average quantiles or prediction intervals, as in \cite{lei:etal:18}. It is sufficient to aggregate the outcomes of individual splits into a final ensemble and then use it, for example, to estimate the selected quantiles. Finally, when the split is performed simultaneously on all the variables analyzed, the obtained residuals maintain the correlation structure of the forecast errors and, therefore, are suitable for approximating the multidimensional distribution.  
We believe that resampling methods are of particular use in this context. They do not require parametric modeling of the multidimensional distribution and are able to approximate well the complex relationship between variables. The multivariate distribution of several variables can then be leveraged in decision-making processes by simultaneously assessing multiple correlated sources of uncertainty or computing functions of the original variables.

The accuracy of the multiple split forecasting approach is evaluated with data describing the German electricity market: DA and ID electricity prices, total load, and RES generation. It is compared with two well-known approaches: quantile regression and historical simulation. The out-of-sample period consists of 730 days, which corresponds to two years of observations. The forecast performance is evaluated with measures that focus on the fit of selected quantiles and explore the distribution of the ensemble. The results lead to the following conclusions: 
\begin{itemize}
    \item QR provides prediction intervals that are less accurate than PIs from ensemble methods (historical simulations and MS). However, the probabilistic forecasts obtained with QR are relatively sharper than those based on other methods, resulting in lower values of the CRPS measure.
    \item The multiple split method allows one to construct prediction intervals that have an accurate empirical coverage. The Kupiec test was unable to reject the null of a correct PI calibration in more than 90\% of the analyzed cases  for $T=365$.
    \item  When different specifications of the MS approach are compared, the model with 20 splits, MS(20), outperforms the MS(1) method. 
    \item Ensemble methods, MS in particular, perform very well when the linear combination of fundamental variables is forecasted. In the case of Price Spread and Residual Load, a multivariate MS(20) approach provides PIs with empirical coverage very close to the nominal one. The Kupiec test confirms the correct empirical coverage in 73.33\% and 92.5\%, respectively. 
    \item When the multidimensional distribution is considered, the ensemble methods that allow for the joint forecasting of the fundamentals provide the best calibration of the probabilistic predictions. Furthermore, the reliability index indicates that the MS(20), \textit{Corr.} approach produces the most accurate fit to the multidimensional distribution. 
\end{itemize}

Next, the economic value is evaluated with an example of a generation utility that owns wind farms spread throughout Germany. The company needs to make bids on the DA market. In particular, it decides on the share of predicted production, $q$, that is offered on the DA market. To balance the position, the remaining part of generation is either sold or purchased on the ID market. The company acts under uncertainty: it knows neither the next-day production nor the future electricity prices. In this research, we consider two benchmark strategies: a naive strategy, which assumes that all predicted generation is offered on the DA market, and a limited bid strategy, which also has $q=1$, but allows for production curtailment when the DA prices fall below zero. These two benchmarks are compared with data-driven approaches that explore the probabilistic prediction of profits. They select the optimal value of $q$ by maximizing the expected profit, the VaR, or Sharpe ratio. Additionally, it is assumed that the company may curtail the generation, as in the limited-bid case. The stopping rule ensures that the utility sells electricity only when a selected quantile, $\tau$, of profits is positive.
The results can be summarized as follows:
\begin{itemize}
    \item The naive benchmark approach is outperformed by all strategies, both in terms of level of profits and risk.
    \item Adopting data-driven strategies without generation curtailment leads to an increase in profits of 1-1.5\% and a decrease in risk. 
    \item Strategies that allow for stoppage of production bring on average higher profits than those without any stopping rule, particularly when profit per trade is considered.
    \item The Limited bid strategy is dominated by data-driven methods for many values of $\tau$. When the quantile of profits used in the stopping rule varies between 0.25 and 0.6 then the Limited bid provides a lower average profit and, at the same time, is characterized by a higher risk than any of the data-driven strategies.
    \item The selection among data-driven strategies depends on the approach to risk. The strategy that aims to maximize expected profits yields the highest income, but at the same time has a lower value of VaR. The opposite can be observed for the method based on VaR. As expected, the approach using the Sharpe ratio balances revenue level and risk.
\end{itemize}

The results show the potential of the MS method in forecasting a multidimensional distribution of variables describing the electricity market and designing trading strategies. They encourage further research on properties and applications that go beyond the presented analysis. First, it would be interesting to develop a rule to choose the number and proportion of splits. Here, we show the results of only two scenarios MS(1) and MS(20) in which these quantities were arbitrarily selected. A more comprehensive analysis in this area is recommended.  Next, since the size of the prediction intervals does not change much over time, one could consider applying locally weighted methods to condition the length of the PI on the fluctuating market situation. This can increase the sharpness of the distribution and improve the statistical performance of the method. Finally, MS can be combined with point forecasting methods other than AR, for example machine learning. Due to a simple construction and flexibility in selecting the number of splits, it can be used together with methods that are computationally burdensome.

\section*{Author contributions}
Conceptualization: K.M., W.N.; Investigation: K.M., W.N.; Software: K.M., W.N.;
Supervision: K.M.; Validation, K.M., W.N. ; Writing - original draft: K.M., W.N. ; Writing - review \& editing: W.N.

\section*{Acknowledgments}
This work was supported by the Ministry of Education and Science (MEiN, Poland) through Diamond Grant No. 0027/DIA/2020/49 (to W.N.) and the National Science Center (NCN, Poland) through SONATA BIS grant No. 2019/34/E/HS4/00060 (to K.M).

\bibliographystyle{unsrtabbr}
\bibliography{bootstrap_lit.bib}

% \printbibliography

\end{document}